\documentclass[fleqn,usenatbib]{mnras}

\usepackage{newtxtext,newtxmath}

\usepackage[T1]{fontenc}

\DeclareRobustCommand{\VAN}[3]{#2}
\let\VANthebibliography\thebibliography
\def\thebibliography{\DeclareRobustCommand{\VAN}[3]{##3}\VANthebibliography}

\usepackage{graphicx}	
\usepackage{amsmath}	
\usepackage{xcolor}
\usepackage{comment}
\usepackage{booktabs}

\usepackage{pdflscape}
\usepackage{longtable}
\usepackage{multirow}

\title[Eclipsing binaries with M-dwarf companions]{Fundamental effective temperature measurements for eclipsing binary stars - IX. Characterisation of 5 solar-type eclipsing binaries with M-dwarf companions.}

\author[A. Hahlin et al.]{
Axel Hahlin$^{1}$\thanks{E-mail: a.j.hahlin@keele.ac.uk}, 
Pierre F. L. Maxted$^{1}$, Ivanna Hern\'andez-Araya$^{2,3}$, Natasha Adshead$^1$, and \newauthor 
Claudia Aguilera-G\'omez$^2$ \\
$^{1}$Astrophysics group, Keele University, Staffordshire, ST5 5BG, UK\\
$^{2}$Instituto de Astrof\'isica, Pontificia Universidad Cat\'olica de Chile, Av. Vicu\~na Mackenna 4860, 782-0436 Macul, Santiago, Chile\\
$^{3}$European Southern Observatory, Alonso de Córdova 3107, Vitacura Santiago, Chile
}

\date{Accepted XXX. Received YYY; in original form ZZZ}

\pubyear{2026}

\begin{document}
\label{firstpage}
\pagerange{\pageref{firstpage}--\pageref{lastpage}}
\maketitle

\begin{abstract}
Large-scale spectroscopic surveys will need suitable benchmark stars that can be used to verify the performance of automatic pipelines. A promising type of object for this purpose is the detached eclipsing binary (DEB) containing an FG-star and a low temperature companion. For flux ratios $\lesssim1$~per~cent, the spectra obtained for the binary will be almost identical to a single star. As the methods for obtaining accurate parameters for binaries are well established, they would be excellent benchmark candidates. We analyse a group of DEBs with M-dwarf companions using TESS light curves and high resolution spectroscopy from NIRPS and HARPS on the ESO 3.6-m telescope. We obtain accurate fundamental parameters for the stellar components. The secondary component is visible with cross-correlation, allowing them to be characterised as well. With the stellar parameters obtained from the analysis, we can measure precise effective temperatures from Gaia parallaxes and archival magnitude data. Chemical compositions of the primary stars are also extracted from the stellar spectra. In this analysis, the secondary components have minimal impact on determined spectroscopic parameters. Our radii and masses are characterised to within sub per~cent precision as well as our primary effective temperatures. Beyond a new set of stellar benchmarks, the secondary components analysed in this work represent a well characterised sample of M-dwarf stars which will be useful to improve our understanding of low mass structure and evolution.

\end{abstract}


\section{Introduction}
Detached eclipsing binaries (DEBs) are some of the best sources of fundamental data of stars. Thanks to developments in space-based photometry and high resolution spectroscopy in the recent decades, the radius and mass of binary components can now be reliably determined with a precision below $\pm0.5$\,per~cent \citep[see e.g.][]{southworth:2015}. One major limitation in using DEBS as sources of fundamental parameters has, up until recently, been the challenge to obtain accurate effective temperatures. One way to obtain effective temperatures of stars is to use distances measured by Gaia \citep{Gaiacollaboration:2016}, in combination with interferometric radius measurements \citep[e.g.][]{huber:2012,rains:2020}. While these measurements have improved the ability to obtain stellar effective temperatures, it has also been shown that interferometric measurements of stellar angular diameters can be subject to significant systematic effects of several per~cent depending on data reduction methods \citep[see e.g.][]{bond:2020}. In the context of determining effective temperatures, systematic errors for angular diameters of a few per~cent would propagate to additional systematic uncertainties in effective temperature of a few 100\,K for Sun-like stars. An alternative to interferometric radius measurements is to use accurate radii determined on eclipsing binaries to obtain stellar angular diameters. Using this approach, it has been possible to measure effective temperatures on eclipsing binaries with a precision of $\pm50$\,K or better on a handful of targets \citep[e.g.][]{miller:2020,miller:2022,maxted:2023}. 

Most DEBS where this precision has been reached have optical flux ratios $L_2/L_1 \approx 1$, i.e. double lined spectroscopic binaries. These systems are beneficial to study due to the ease of performing radial velocity measurements of both components simultaneously. On the other hand, the significant contribution of both stars to the signal makes them challenging to use for verification of automatic pipelines that measure atmospheric parameters of stars, as their spectra will be substantially different from the typical single star. Instead, DEBS with small optical flux ratios $L_2/L_1 \lesssim 0.01$, also known as low-mass eclipsing binaries \citep[EBLMs,][]{triaud:2013}, could be used for this verification. Using near-infrared (NIR) observations, where the flux ratio will be higher, the secondary component can still be characterised by using multi-line techniques such as cross-correlation as demonstrated by \cite{maxted:2023}. The result will be systems that observationally looks almost identical to a single star, but with the benefits in fundamental parameter precision provided by DEBS. These systems would be suitable as benchmark stars for missions such as PLATO \citep{rauer:2014}. Furthermore, if the spectral type of the primary is F or G, the secondary component will be an M-dwarf (or possibly a late-K dwarf). This means that the secondary component will not interfere with asteroseismology measurements of the system during the PLATO mission, as M-dwarfs do not exhibit currently detectable asteroseismic variability \citep[e.g.][]{rodriguez:2016}. 

\cite{maxted:2023} outlined some properties that would be beneficial for these benchmark systems. As already mentioned, the system should have a flux ratio of $\lesssim1$\,per~cent and have a FG-type primary. The system should also have magnitudes in the range $V=9$--$12$. While this range is $\sim5$ magnitudes lower compared to typical benchmark stars, it means that they will have similar magnitudes to the stars that are commonly targeted by large-scale spectroscopic surveys such as APOGEE \citep{majewski:2017}, and upcoming surveys such as 4MOST \citep{dejong:2019} and PLATO. The separation of the components should also be sufficient to avoid interactions such as tidal distortions and rapid rotation due to synchronisation. For this purpose, \cite{maxted:2023} provided a cut-off at orbital periods $P<4$\,days. To avoid very active stars, the out of eclipse variations due to starspots should also be low. Still, some stars with moderate magnetic activity, showing starspot variations of $\approx 1$\,per\,cent, are included. While activity may not be ideal for a benchmark stars, many stars do exhibit activity and understanding the impact of magnetic activity is important for the reliability of future surveys.

Beyond using the primary components as benchmark stars, the accurate characterisation of the secondary components would also be useful. Understanding processes such as radius inflation that is commonly observed on low-mass stars \citep[e.g.][]{casagrande:2008,spada:2013} will require accurate parameters \citep{sebastian:2023}. This inflation has been tied to magnetic activity \citep{feiden:2013,kesseli:2018}, with support from direct measurements of surface magnetic fields on M-dwarfs \citep{kochukhov:2019,hahlin:2024}. Starspots, another manifestation of stellar activity, have also been used to explain radius inflation on low-mass stars \citep[e.g.][]{cao:2025}. Another challenge with EBLM systems is to study tidal interactions of the secondary component. While studies on the primary components of EBLM systems have been carried out by \cite{sethi:2026}, improved information on the secondary could help our understanding of how the M-dwarf companions are affected by tidal interactions in EBLM systems.

In this work, we analyse five DEBS that satisfy the criteria above. Some were already listed by \cite{maxted:2023} (CD-31 3271, HD 4875, TYC 8547-22-1) as FG-type primaries with M-dwarf companions that would be suitable benchmark stars. Some of those stars had dynamical velocities from Gaia, allowing for preliminary mass and radius estimations. We also include two binaries that have similar properties (HD 287990, BD-08 1175) but that were not included in \cite{maxted:2023}. A compilation of basic properties can be seen in Table~\ref{tab:basic_info}. In Sect.~\ref{sec:OBS} we cover the spectroscopic and photometric observations that were obtained for the analysis. We cover the analysis that was carried out for the stars of this sample in Sect.~\ref{sec:METHODS}. Any specific detail in the analysis for each star is presented in Sect.~\ref{sec:individual}. In Sect.~\ref{sec:discussion}, we use the obtained properties to discuss the extent of tidal interaction and how well predicted stellar age matches the observed stellar structures. Finally, the results are summarised in Sect.~\ref{sec:conc}. 

\section{Observations}
\label{sec:OBS}

\begin{table*}
	\centering
	\caption{Parameters of the eclipsing binaries investigated in this work. Spectral types (Sp.) are taken from Simbad or estimated based on T$_{\rm eff}$. $N_{*}$ indicates the number of nights each target was observed, or the number of eclipses detected in the TESS light curves.}
\label{tab:basic_info}
\begin{tabular}{lccrlrrc} 
\hline
Name & $\alpha$ (J2000.0) & $\delta$ (J2000.0) & $G$ [mag] & Sp. & $N_{\rm NIRPS}$ & $N_{\rm HARPS}$ & $N_{\rm Eclipse}^*$
\\
		\hline
CD$-$31 3271   & 06:24:48.93 & $-$31:51:52.2 &  9.81 & G0  & 11 & 7 & $7 + 8$ \\
HD 4875        & 00:50:39.98 & $-$18:30:21.2 &  8.82 & G3V & 15 & 2 & $1+2$  \\
HD 287990      & 05:31:04.20 & $+$01:11:15.5 &  9.90 & F5  &  7 & 3 & $1+2$ \\

TYC 8547-22-1  & 06:36:58.94 & $-$58:27:36.6 &  9.96 & G0  &  8 & 8 & $221 + 216$  \\
BD-08 1175     & 05:36:55.82 & $-$08:47:57.6 & 10.12 & G0  &  9 & 7 & $2+4$\\
\hline

\noalign{\smallskip}
\multicolumn{8}{l}{{\bf Notes:} $^{*}$ Primary and secondary eclipses shown separately.}
\end{tabular}
\end{table*}

\subsection{NIRPS spectroscopy}
\label{sec:NIRPS}
We observed a collection of eclipsing binaries during the ESO observing period P112 using the high resolution spectrograph NIRPS \citep{wildi:2017,bouchy:2017}, mounted on the ESO 3.6-m telescope. The spectrograph covers the NIR wavelengths between 970 and 1850~nm (the Y, J, and H band). For our observations, we used the high efficiency observing mode with a resolving power of $R\approx70\,000$.

To obtain dynamical information, each target was observed on multiple nights at different orbital phases. In addition, some targets were observed multiple times in sequence to ensure sufficient S/N could be achieved. The combined spectrum of each night typically had S/N ranging between 50 and 100. After the observations have been carried out, we obtain the reduced and telluric corrected data from the ESO science archive\footnote{\url{https://archive.eso.org/scienceportal/home}}. 

We then proceed with a few extra reduction steps before beginning the main scientific analysis. While the NIRPS spectra have been corrected for telluric contamination, strong tellurics can still produce significant artefacts in the spectra. For this reason, we remove the data in-between the Y-, J-, and H-bands. We then continuum normalise the spectra using splines for each band separately. Finally, for the targets with multiple consecutive observations, we stack the spectra observed each night together. As the spectra observed for the same star on any given night were carried out in sequence, no significant radial velocity shift could be detected between individual spectra of each night. For this reason, they were stacked without any additional radial velocity shifts. The number of nights each star was observed is shown in Table~\ref{tab:basic_info}.

\subsection{HARPS Spectroscopy}
\label{sec:HARPS}
We also obtained spectra with the HARPS instrument \citep{mayor:2003} on the ESO 3.6-m telescope simultaneously with the NIRPS spectra described in Sect.~\ref{sec:NIRPS}. HARPS is a high resolution spectrograph covering optical wavelengths between 380 and 690\,nm, the instrument has a blue arm (378--530\,nm) and a red arm (533-691\,nm). We use the high efficiency mode with a resolving power  $R\approx 80\,000$ for our observations. We downloaded the reduced data from the ESO science archive. For each spectrum, we normalise the individual arms separately using a 5th order polynomial. Similar to the NIRPS data, we combine any observation obtained on the same night. The total number of HARPS observation nights are shown in Table~\ref{tab:basic_info}.

\subsection{Photometry}
\label{sec:photometry}
We utilise light curves obtained from the TESS mission \citep{ricker:2015}. Our objects have multiple eclipses visible in the light curves, ranging from a few to several hundred (see Table~\ref{tab:basic_info}). We used {\sc lightkurve}\footnote{\url{https://docs.lightkurve.org/}}  \citep{lightkurvecollaboration:2018} to search the Mikulski Archive for Space Telescopes\footnote{\url{https://archive.stsci.edu/}} (MAST) for light curves of each binary system observed by the TESS mission. We used light curves with a cadence of 120\,s and processed to produce {\sc pdc\_sapflux} values by the TESS Science Processing Operations Center (SPOC). The data were downloaded from MAST using {\sc lightkurve} and bad data were rejected using the default bit mask. We only included the data that was observed during or close to the primary or secondary eclipses. Each segment (data during and surrounding each eclipse) were normalised by fitting a straight line to the light curve regions before and after each eclipse.

\section{Methods}
\label{sec:METHODS}
The procedures we used to measure the properties of the selected binary stars from their light curves and spectra are described in this section. Many of these methods are dependent on stellar parameters obtained in later analysis steps. We iterated the analysis, starting from literature values, until the resulting stellar parameters stabilised. We found that we did not need more than two iterations for any star in our sample, indicating that the approach quickly finds a stable solution. Variations in these procedures for individual systems are described in Section~\ref{sec:notes}.

\subsection{Spectroscopic analysis - NIRPS determination of secondary parameters}
\label{sec:spectroscopy}
Our analysis is based on the method outlined in \cite{adshead:2026}. The primary difference is that the reduced data available on the ESO science archive is provided as a single segment, rather than divided into orders. For this reason, our cross-correlation functions (CCF) have been determined using the data as a single segment. To verify that this did not introduce any systematic differences from \cite{adshead:2026} we also investigated CD-27 2812. We find that our value of $K_2$ is consistent with the value found by \citeauthor{adshead:2026}. As such, we do not expect this difference in the analysis to impact our results significantly. 

From the spectra stacked by night as described in Sect.~\ref{sec:NIRPS} and \ref{sec:HARPS} we first determine the radial velocities of the primary component using cross-correlation with \textsc{iSpec} \citep{blancocuaresma:2014,2019MNRAS.486.2075B}. We use a line mask corresponding to solar parameters for all of our primary stars. The obtained radial velocities are presented in Table~\ref{tab:rvs}. From the obtained radial velocities, together with the light curves, we determine orbital parameters of the system (see Sect.~\ref{sec:light curve}). To verify that there is no systematic shift between the two datasets, we also fit the systemic velocity, $V_0$, and the primary semi-amplitude, $K_1$, using only radial velocity data, this produces no significant change in dynamical parameters of the primary. 

\begin{figure*}
    \centering
    \includegraphics[width=\linewidth]{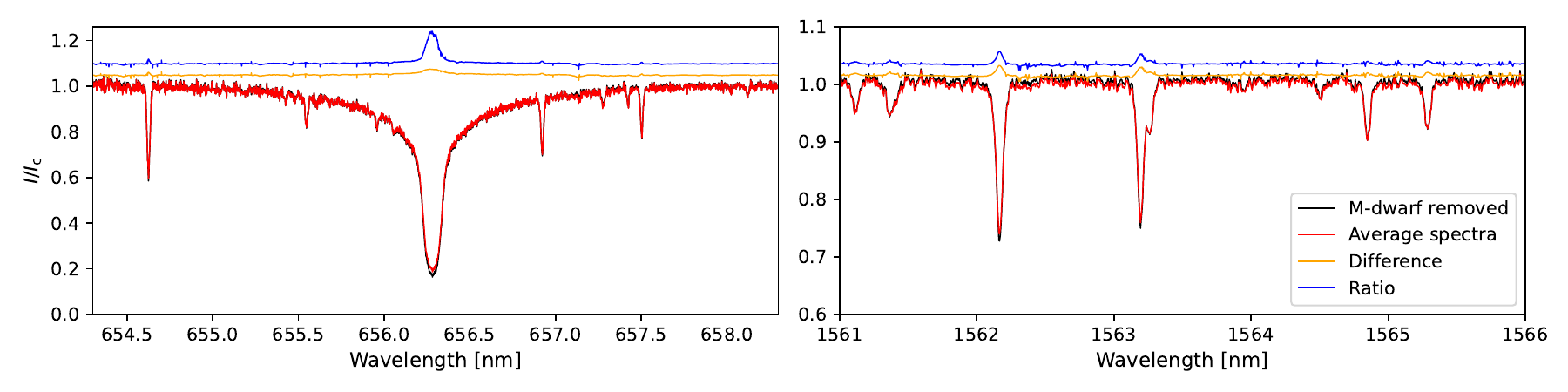}
    \caption{Segments of the mean spectra of BD-08~1175 from both HARPS (left) and NIRPS (right). Shown is both the spectra obtained from stacking ($S_0$), and the spectra with the M-dwarf component removed ($S_{\rm -M}$). The difference ($S_0-S_{\rm -M}$) and ratio ($S_{0}/S_{\rm -M}$), shifted for visibility, are also shown.}
    \label{fig:M_removal}
\end{figure*}

To investigate the signal of the secondary component, we use the obtained radial velocities of the primary to produce a mean spectrum of the primary star. This is done by shifting all spectra into the rest frame of the primary star and stacking them together with a weighted mean. We also perform an outlier removal in the stellar rest frame to ensure that any telluric or interstellar absorption still present in individual spectra have a minimal effect on the mean spectra. A segment of the mean spectra can be seen in Fig.~\ref{fig:M_removal}. We then divide the spectrum of each night with the obtained mean spectrum. This produces a series of spectra with the primary contribution removed. To determine the secondary semi-amplitudes, $K_2$, of the secondary components, we perform cross-correlation with a synthetic M dwarf spectra from the BT-Nextgen grid \citep{allard:2011} obtained from the Spanish virtual observatory\footnote{\url{https://svo2.cab.inta-csic.es/theory/newov2/}}. The initial synthetic spectra use expected parameters of the secondary taken from \cite{maxted:2023}. Once the stellar parameters are more constrained, we repeat the cross-correlation with an updated mask. We find that the choice of mask is not significant for the determination of $K_2$ as long as it is in reasonable agreement with the M-dwarf parameters. We also apply a low frequency noise filter to the data. The CCF obtained from each night is then stacked into an average CCF. To find the $K_2$-value of the secondary, we perform a Gaussian fit to the profile in combination with Gaussian processes using {\sc celerite} \citep{foremanmackey:2017} to model the variability in the CCF not caused by the secondary. To obtain posterior distributions of the orbital parameters of the secondary, we also perform MCMC-sampling using \textsc{emcee} \citep{foremanmackey:2013}.

Once the $K_2$-value has been determined, we refine the mean spectrum of the primary by removing the contribution of the secondary from each spectrum. As the flux ratio contribution of the secondary is $\lesssim$ few per~cent for our targets, we remove the secondary contribution by using synthetic models from the BT-Nextgen grid convolved with a Gaussian profile to obtain the same resolution as the observed spectra. A comparison between the two mean spectra can be seen in Fig.~\ref{fig:M_removal}. While there is some dilution of line depths, the effect is generally quite small and no spectral signals from the M-dwarfs can be seen. This is likely since stacking multiple spectra smears out the spectroscopic signal of the M-dwarf. We then repeat the determination of $K_2$. We find that this has no significant impact on our obtained $K_2$ values. The M-dwarf templates were generated using the stellar parameters iteratively obtained in the analysis. Due to the small flux ratios, the choice of M-dwarf parameters has a small impact on the resulting primary spectra provided that the model effective temperature is within a few 100\,K of the actual M-dwarf effective temperature.

\begin{figure}
    \centering
    \includegraphics[width=\linewidth]{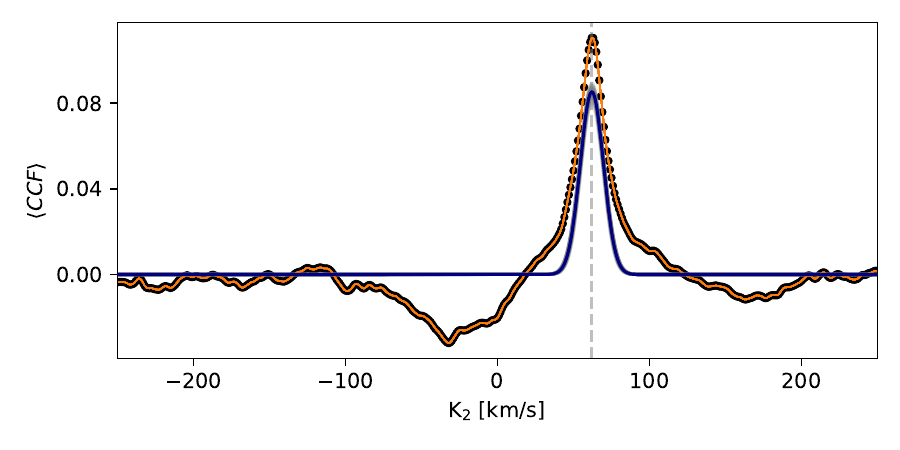}
    \caption{Mean CCF of BD-08~1175~B (black) along with the best-fit (orange). The Gaussian component of the fit is shown independently (blue), the gray region indicates its variation by randomly plotting 50 MCMC samples.}
    \label{fig:CCF_example}
\end{figure}

To test the stability of our method, we measure the $K_2$-value using four different wavelength ranges. The entire NIRPS range, the Y-band, the J-band, and the H-band. While all bands agree reasonably well, it gives another indication of the uncertainty in our $K_2$ determination. From this investigation, we found that the H-band is the dominant contribution to the CCF obtained from the full spectrum. This is likely because the flux ratio between the two components should be larger at longer wavelengths, which results in a stronger CCF peak for the secondary component. In addition, removing the M dwarf contribution from the primary spectra reduces the variation caused by the choice of wavelength region the CCF is determined over. In the end, we find that the standard deviations between the different bands are typically around 0.1 -- 0.2~km~s$^{-1}$, which is generally comparable to our resulting uncertainties shown in Table~\ref{tab:lcfits}. A resulting CCF, averaged over all observing nights, and its best fit can be seen in Fig.~\ref{fig:CCF_example}. We note that the discrepancy between wavelength bands is dependent on the cross-correlation mask. For larger discrepancies in T$_{\rm eff}$ of around 500\,K, we found that the mask could induce variations of $K_2$ of ~1\,km\,s$^{-1}$. However, the choice of line mask primarily affected the Y- and J-band while the $K_2$ values from the H-band and the whole spectrum remained relatively stable. Regardless, using a consistent mask when determining $K_2$ is important to minimize systematic effects.

\subsection{Spectroscopic analysis - Chemical composition from HARPS spectra}
\label{sec:optical_spectroscopy}

We investigate the chemical composition of the binaries by using the observations from HARPS described in Sect.~\ref{sec:HARPS}. Observations from each night is combined into average spectra of each target following the same method as for the NIRPS spectra described in Sect.~\ref{sec:spectroscopy}. Each spectrum is shifted to the same wavelength using the radial velocities obtained from cross-correlation of the HARPS spectra with a line mask corresponding to solar parameters. 

To test the impact of the secondary component, we use the dynamical information of the secondary obtained in Sect.~\ref{sec:spectroscopy} to subtract out the M dwarf component from the individual spectra using the same BT-Nextgen grid of synthetic spectra as described in Sect.~\ref{sec:spectroscopy}. A segment around the H$\alpha$-line of the resulting HARPS mean spectra, both with and without the M-dwarf signal removed, can be seen in Fig.~\ref{fig:M_removal}. The effect is marginal, but clearly noticeable in strong lines such as H$\alpha$.

For the chemical analysis, we used both versions of the spectrum: the original spectrum, which contains the flux contribution from both components of the binary system, and the M-dwarf-corrected spectrum, with the contribution of the secondary companion removed. The goal of analysing both spectra was to determine whether the secondary component has a measurable impact on chemical abundances of the primary star. Therefore, we applied the same homogeneous spectroscopic analysis to both sets of spectra.

The analysis was performed with iSpec, a software framework designed for the treatment and analysis of stellar spectra, including the determination of atmospheric parameters and chemical abundances. We used spectral synthesis together with the radiative transfer code MOOG (\citealp{1973PhDT.......180S}) and MARCS model (\citealp{2008A&A...486..951G}) atmospheres. The adopted line list was the GES v6 (\citealp{2021A&A...645A.106H}), line list prepared for MOOG within iSpec, and the solar abundances were taken from \cite{2007SSRv..130..105G}.

\begin{figure}
    \centering
    \includegraphics[width=1.\linewidth]{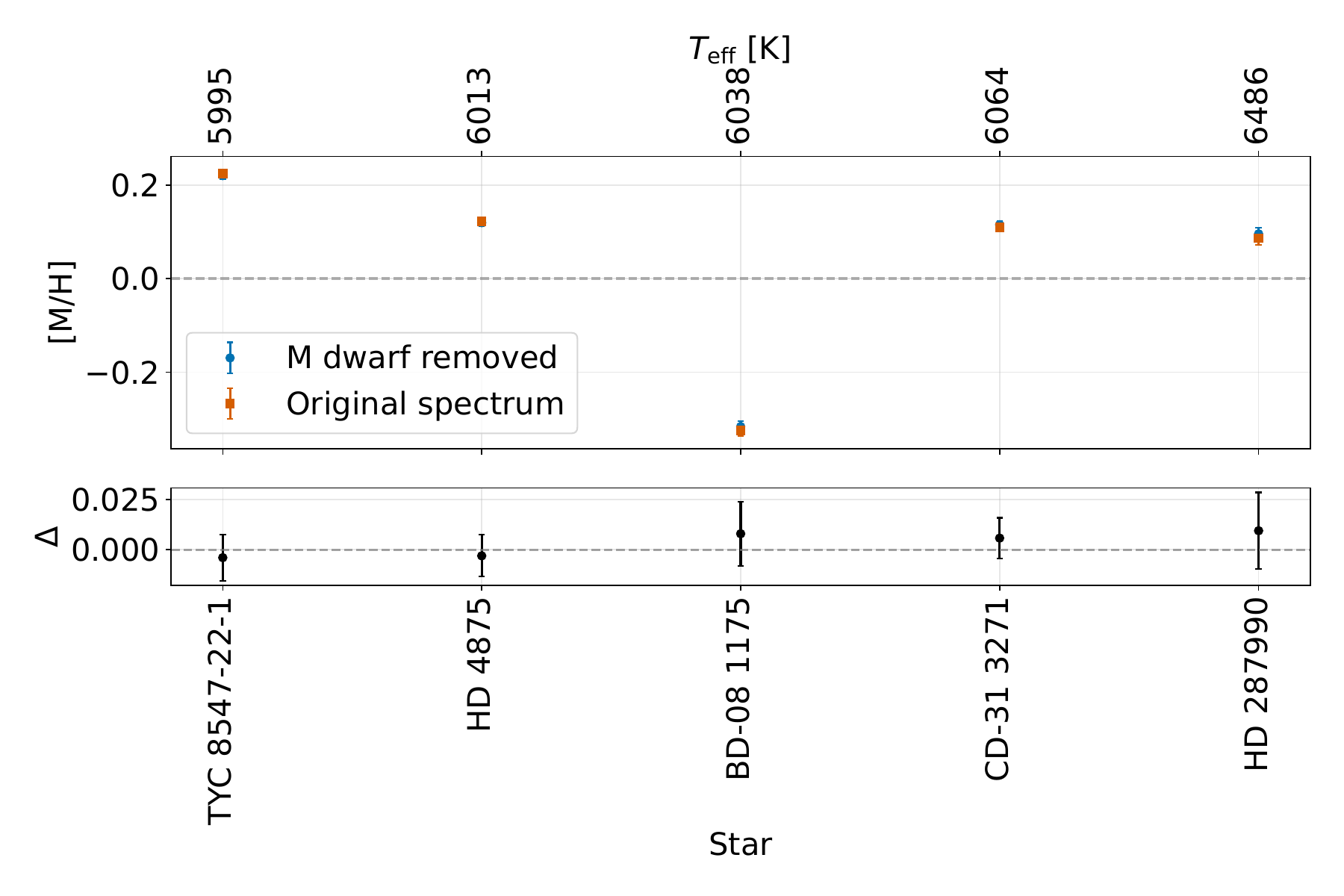}
   \caption{Comparison of the global metallicity, $\mathrm{[M/H]}$, derived from the original spectra and from the M-dwarf corrected spectra. The upper panel shows the values obtained for each target, with the corresponding $T_{\rm eff}$ indicated on the upper x-axis. The lower panel shows the difference $\Delta\mathrm{[M/H]} = \mathrm{[M/H]}_{\rm M-dwarf,corrected} - \mathrm{[M/H]}_{\rm original}$. Error bars correspond to the formal uncertainties returned by iSpec.}
    \label{fig:mh}
\end{figure}

After selecting this configuration, we derived the final global metallicity by fixing $T_{\rm eff}$ and $\log g$ to the reference values obtained from the fundamental analysis of the systems (see Sect.~\ref{sec:fundamental_par}, \ref{sec:teb}, and Table~\ref{tab:massradius}), while leaving metallicity [M/H], microturbulent velocity $v_{\rm mic}$, and projected rotational velocity $v\sin i$ as free parameters. The macroturbulence velocity was calculated using an empirical relation presented by \cite{2014A&A...566A..98B}, suitable for FGK spectral type stars. The resulting global metallicities are reported in Table~\ref{tab:metallicity_comparison}. The differences between the metallicities obtained from the original spectra and the M-dwarf corrected spectra are shown in Fig.~\ref{fig:mh}. We find that the differences in [M/H] are always smaller than 0.01 dex. The mean absolute difference is only $\sim 0.006$ dex, indicating that the contribution of the M-dwarf companion has a negligible effect on the global metallicity derived from the optical HARPS spectra.

\begin{table}
    \centering
    \caption{Comparison of the metallicities derived from the original spectra and from the spectra after removing the M-dwarf contribution. The last column shows the difference between both. The uncertainties of the individual metallicities are those returned by iSpec.}
    \label{tab:metallicity_comparison}
    \begin{tabular}{lccc}
    \toprule
    Star 
    & ${\rm [M/H]}_{\rm -M}$
    & ${\rm [M/H]}_{\rm 0}$ 
    & $\Delta{\rm [M/H]}$ \\
    & dex & dex & $\sigma$ \\
    \midrule
    CD$-$31 3271 & $+0.115 \pm 0.007$ & $+0.109 \pm 0.007$ & $+0.61$ \\ 
    HD 4875 & $+0.119 \pm 0.007$ & $+0.122 \pm 0.007$ & $-0.30$\\
    HD 287990 & $+0.096 \pm 0.013$ & $+0.086 \pm 0.014$ & $+0.52$ \\ 
    TYC 8547$-$22$-$1 & $+0.220 \pm 0.008$ & $+0.224 \pm 0.008$ & $-0.35$\\ 
    BD$-$08 1175 & $-0.316 \pm 0.011$ & $-0.324 \pm 0.011$ & $+0.51$\\ 
    \bottomrule
    \end{tabular}
    
\end{table}

\begin{figure}
    \centering
    \includegraphics[width=1\linewidth]{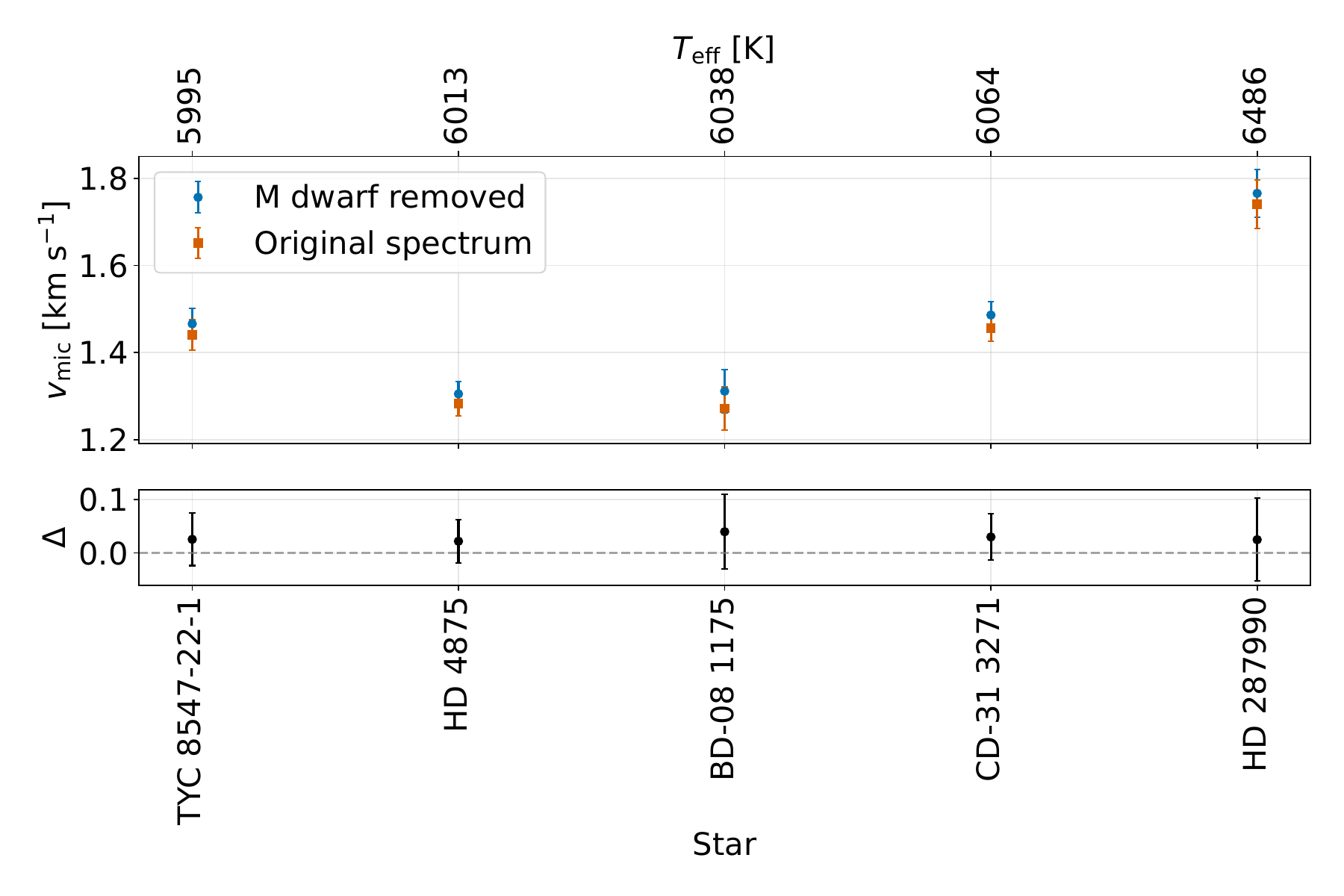}
    \caption{
    Microturbulent velocity, $v_{\rm mic}$, obtained from the original and M-dwarf corrected spectra. The upper panel shows the values for each target, while the lower panel shows $\Delta v_{\rm mic} = (v_{\rm mic})_{\rm M-dwarf,corrected} - (v{\rm mic})_{\rm original}$. Error bars correspond to the formal uncertainties returned by iSpec.
    }
    \label{fig:vmic}
\end{figure}

\begin{figure}
    \centering
    \includegraphics[width=1.\linewidth]{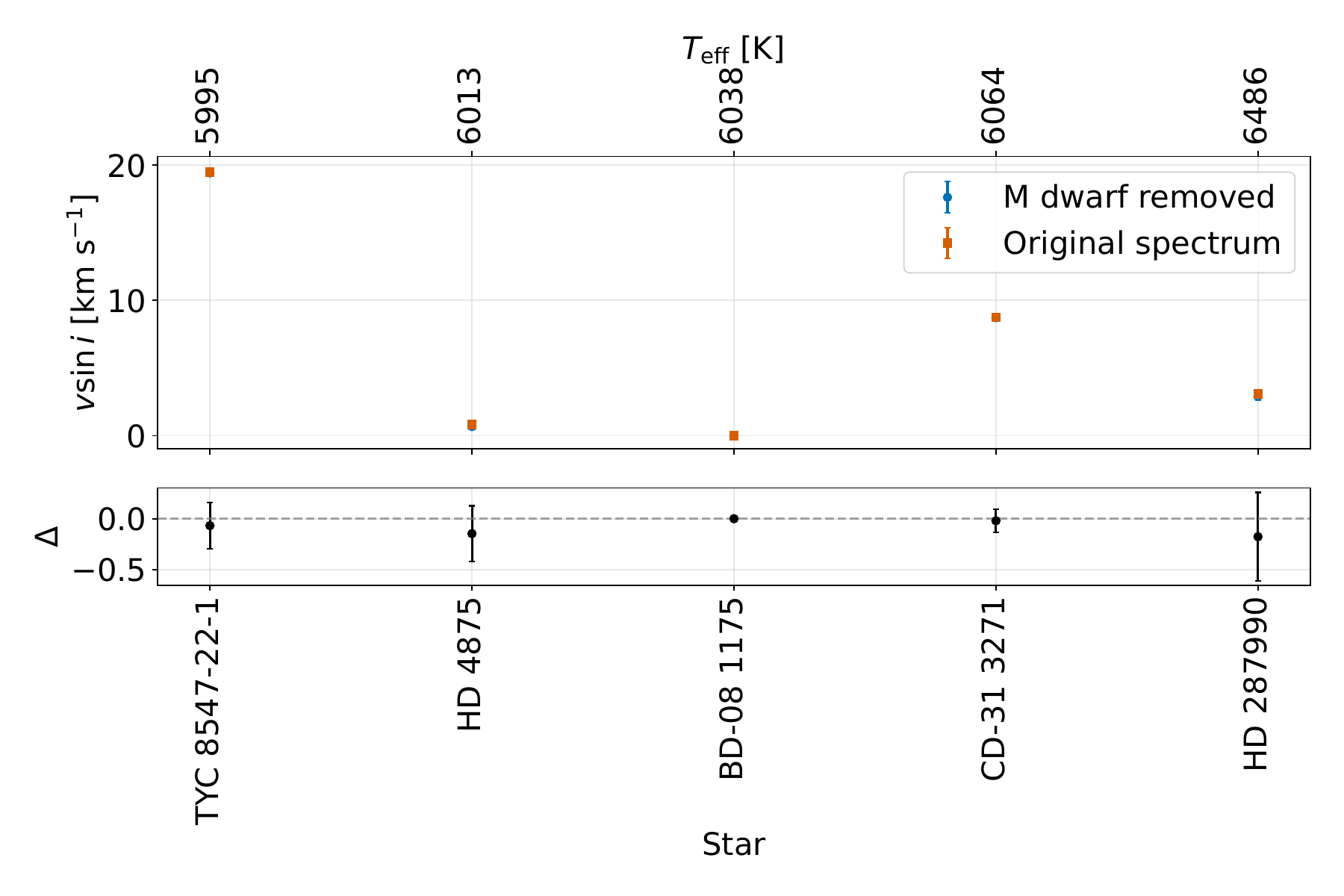}
    \caption{Projected rotational velocity, $v\sin i$, obtained from the original and M-dwarf corrected spectra. The upper panel shows the values for each target, while the lower panel shows $\Delta v\sin i = (v\sin i)_{\rm M-dwarf,corrected} - (v\sin i)_{\rm original}$. Error bars correspond to the formal uncertainties returned by iSpec.}
    \label{fig:vsini}
\end{figure}

\begin{figure*}
    \centering
    \includegraphics[width=1.\linewidth]{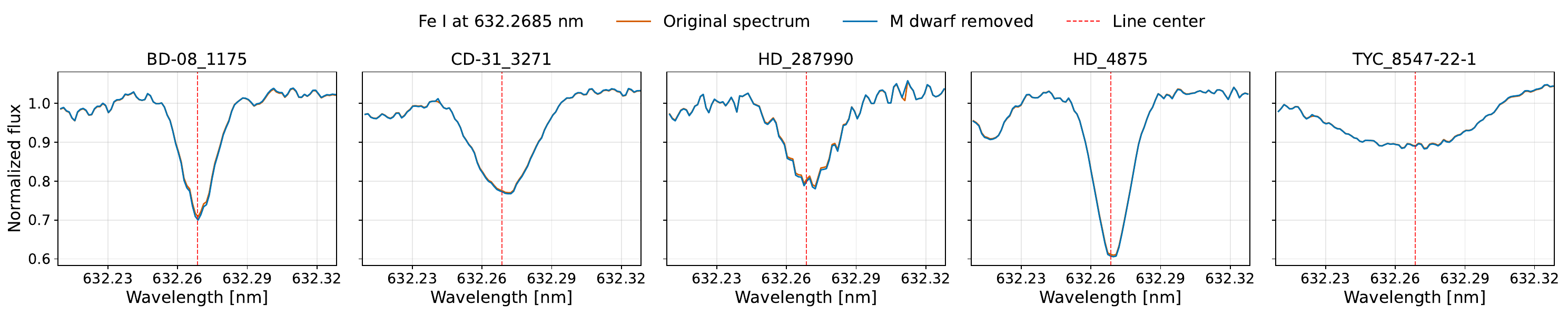}
    \caption{Comparison of the Fe I line at 632.2685 nm for the five targets. In each panel, the original spectrum is shown in orange and the M-dwarf corrected spectrum in blue. The vertical dashed red line indicates the line Fe I.}
    \label{fig:Fe_1_632}
\end{figure*}

Nevertheless, small differences are found in other fitted parameters, particularly in $v_{\rm mic}$ and $v\sin i$. These differences suggest that the M-dwarf contribution, although small, can slightly affect the line profiles and therefore the broadening parameters. As shown in Fig.~\ref{fig:vmic}, the $v_{\rm mic}$, changes by only $\sim 0.02$--$0.04$ km s$^{-1}$. Similarly, the differences in $v\sin i$ are small, with the largest difference being $\sim 0.18$ km s$^{-1}$ for HD 287990 (Fig.~\ref{fig:vsini}). These variations are comparable to, or smaller than, the formal uncertainties of the fit. Therefore, the comparison between the two spectra suggests that the removal of the M-dwarf contribution does not significantly affect the derived stellar parameters. These small differences in obtained stellar parameters are consistent with the visual comparison between the two spectra, as shown in Fig.~\ref{fig:Fe_1_632}.

TYC 8547-22-1 and CD-31 3271 show the largest broadening in the sample, with $v\sin i \simeq 19.4$ km s$^{-1}$ and $v\sin i \simeq 8.7$ km s$^{-1}$, respectively.  This can also be seen in Fig.~\ref{fig:Fe_1_632}, the absorption lines of TYC 8547-22-1 appear broader than those of other stars in the sample. Some care should be taking when considering these $v\sin i$ values as its effect on the stellar line shape is similar to other broadening effects such as macroturbulence and resolution. These effects are therefore challenging to disentangle \citep{2019MNRAS.486.2075B}. It is likely that our empirical macroturbulent velocities from \cite{2014A&A...566A..98B}, influence the obtained $v\sin i$ values, particularly for our slowly rotating stars where the macroturbulent velocity is the dominant broadening contribution.

After deriving the global metallicity and the final atmospheric solution, we used the same spectroscopic setup to determine individual chemical abundances. In this step, all stellar parameters were fixed to the final values obtained previously, and only the abundance of each element was allowed to vary. We used a different line list optimized for abundance determination from \cite{Hernandex-Araya}, allowing us to measure elements belonging mainly to two groups: $\alpha$ elements and iron-peak elements. The resulting chemical abundances for each star are shown in Table~\ref{tab:gbsv3_abundances}.

The abundance analysis was performed independently for the original spectra and for the M-dwarf corrected spectra, following exactly the same procedure in both cases. This allows us to evaluate whether the removal of the M-dwarf contribution has any significant impact.

We compared our derived abundances with those reported for the Gaia Benchmark Stars by \citealp{2026A&A...705A.167C}, adopting their MOOG results for consistency. This comparison is shown in Fig.~\ref{gbsv3_abundances}, where $\mathrm{[X/Fe]}$ is plotted as a function of $\mathrm{[Fe/H]}$ and the points are colour-coded by effective temperature. The reference GBS v3 abundances are shown as triangles, while our results from the original spectra and the M-dwarf corrected spectra are shown as squares and circles, respectively. In general, the two sets of results almost overlap for most elements, indicating that the M-dwarf correction has a negligible impact on the derived abundances. Both sets are also in good agreement with the reference abundance trends.

The main exception to this general agreement is Ti\,I, for which our abundances are systematically lower than the GBS v3 reference values. This offset is present in both the original and the M-dwarf corrected spectra, suggesting that it is not caused by the subtraction of the secondary component. In contrast, Ti\,II shows a much better agreement with the reference abundances. This suggests that the discrepancy is specific to the Ti\,I lines and may be related to line selection or blending effects.

\subsection{Light curve analysis}  
\label{sec:light curve}
We use the term ``eclipse'' to refer to both the primary (deeper) eclipse caused by the transit of the primary star by the secondary star, and the secondary eclipse caused by the occultation of the cooler, smaller secondary star. To analyse the systems, we used the light curves from TESS described in Sect.~\ref{sec:photometry} as well as the radial velocities obtained from the HARPS and NIRPS spectra described in Sect.~\ref{sec:spectroscopy}.
 
To obtain orbital parameters for each system, we used {\sc jktebop}\footnote{\url{http://www.astro.keele.ac.uk/jkt/codes/jktebop.html}} \citep{Southworth:2010}, which is built on the NDE light curve model \citep{nelson:1972}. We approximate that the stars are spherical. As the targets of this study are detached eclipsing binaries with periods $>4$\,days, the systems show no significant signs of oblateness. We also ignore the reflection effect on each system, as well as the ellipsoidal effect. We use the power-2 law from \cite{southworth:2023} to model the stellar limb darkening. This law uses the parameters $h_1$, the relative intensity at the limb angle 0.5, and $h_2$, the difference in relative intensity between limb angles 0.5 and 0. Due to the significant difference in flux contribution, it is primarily the limb darkening of the primary that affects the shape of the eclipses. For this reason, we only let $h_{1,A}$ be a free parameter while $h_{2,A}$ is fixed based on \cite{maxted:2018} at values around 0.4. For the secondary, we fix $h_{1,B}$, and $h_{2,B}$ to 0.7 and 0.4. These values are also based on \citep{maxted:2018} for the TESS bandpass and at the lowest effective temperatures available in that sample. While this is hotter than some of the M-dwarfs in our sample, the low flux contribution means that choice of limb darkening has negligible effect on the light curve fitting.

We use the following parameters for the binary model: the sum of the stellar radii in units of the semi-major axis (fractional radii), $r_1+r_2=(R_1+R_2)/a$; the ratio of the stellar radii, $k=R_2/R_1$; the surface brightness ratio at the disk centre, $J_0$; the orbital inclination, $i$; the time of mid-primary eclipse, $T_0$; the orbital period, $P$;  $e\sin(\omega)$ and $e\cos(\omega)$, where $e$ is the orbital eccentricity and $\omega$ is the longitude of periastron for the primary star. If results show that both $e\sin(\omega)$ and $e\cos(\omega)$ are consistent with 0, we keep them fixed at zero for the subsequent fit. We also investigate the possibility of contamination by allowing the third light of the system to be non-zero if it shows a $1\sigma$ deviation from zero. 

As we have radial velocity measurements of the primary from Sect.~\ref{sec:spectroscopy} and \ref{sec:optical_spectroscopy}, $K_1$ and $V_0$ are included as free parameters in the fit. In some cases, we found that there was a systematic offset between the radial velocity measurements obtained from HARPS and NIRPS. To mitigate this, we allowed $V_0$ to be fit independently for the two spectrographs when a systematic offset is detected. This offset was $\lesssim0.1$~km~s$^{-1}$ for all systems and would show up as an offset in $V_0$ rather than affecting our orbital parameters.

In \cite{maxted:2023}, the mid-eclipse for the primaries and the orbital periods were determined with a combination of observations from TESS and WASP. The motivation was to increase the baseline for improved precision. While we do not use the WASP data to expand the baseline, we do have access to additional observations (radial velocities and longer TESS baselines). However, if our uncertainties in $P$ and $T_0$ are larger than in \cite{maxted:2023}, we fix the period to the value presented in \citep{maxted:2023}. Testing the impact of this shows that this has no significant influence on the fundamental parameters of our sample.
 
\begin{figure}
    \centering
    \includegraphics[width=\linewidth]{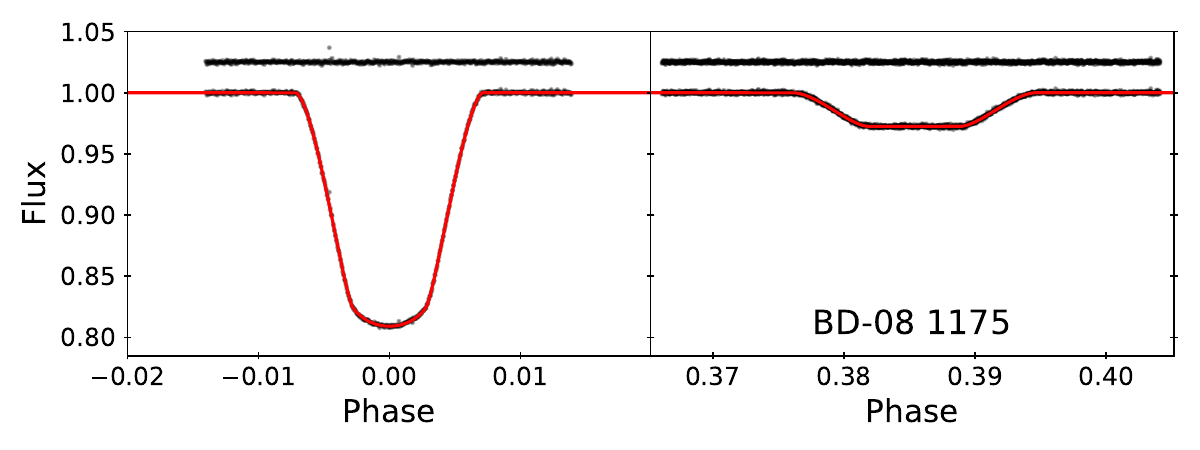}
    \caption{Phase-folded TESS light curves of BD-08~1175 shown as black dots with the best fit model from {\sc jktebop} shown as a red line. The residuals to the light curve is shown above the main eclipses. \textit{Left.} Primary eclipse, \textit{Right.} Secondary eclipse}
    \label{fig:lc_fit}
\end{figure}

\begin{figure}
    \centering
    \includegraphics[width=\linewidth]{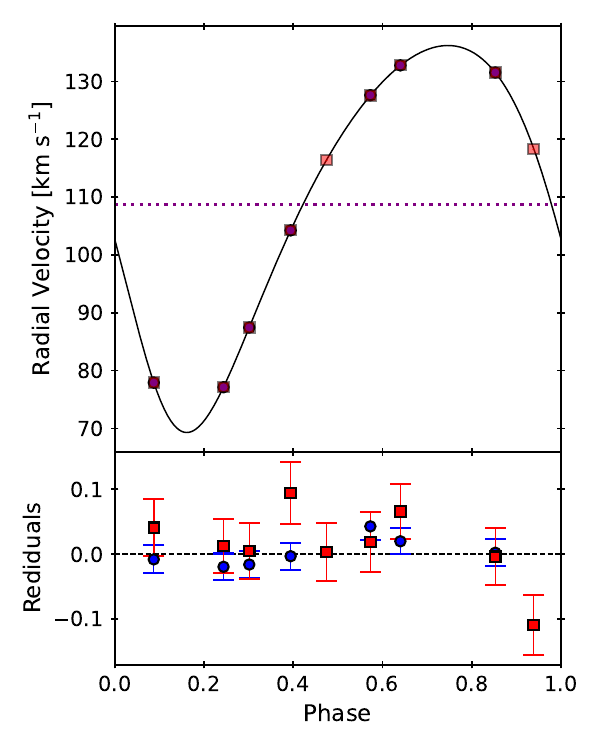}
    \caption{\textit{Top.} Black line shows the fit to the radial velocities of BD-08~1175~A measured from HARPS (blue circles) and NIRPS (red squares) spectra. The NIRPS data has been shifted with an offset of $-0.065$~km~s$^{-1}$ as discussed in Sect.~\ref{sec:light curve}. The systemic velocity $V_0$ is shown as a dotted line. \textit{Bottom.} Residuals to the fit.}
    \label{fig:rv_fit}
\end{figure}

Example fits to the TESS light curve and spectroscopic radial velocities for BD-08~1175 are shown in Fig.~\ref{fig:lc_fit} and~\ref{fig:rv_fit}, respectively. The best-fit parameters for the fit of each star are given in Table~\ref{tab:lcfits}. The values in Table~\ref{tab:lcfits} are the mean and standard error of the mean of the best-fit parameters. The standard error estimates in Table~\ref{tab:lcfits} were estimated using a Monte Carlo method with 1000 independent trials fitting synthetic data generated from the best-fit light curve model. To account for systematic effects of the observation, errors were scaled based on the fit quality. Specifically, Gaussian noise with the same standard deviation as the root-mean-square ({\it rms}) of the residuals, were included in the sampling. 

\begin{table*}

\centering
\caption{Best-fit parameters from the fits to TESS light curves and radial velocities from HARPS and NIRPS. The value in the column $\ell$ is the flux ratio in the TESS band. Figures in parentheses are the standard error on the final digit of the preceding value. }
\label{tab:lcfits}
\begin{tabular}{lrrrrrrr} 
\hline
Name &  
\multicolumn{1}{c}{$P$ [d]} &
\multicolumn{1}{c}{$T_0$ [BJD$-2450000$]}&
\multicolumn{1}{c}{$J_0$}& 
\multicolumn{1}{c}{$r_1+r_2$}& 
\multicolumn{1}{c}{$k = r_2/r_1$}& 
\multicolumn{1}{c}{$h_1$}& 
\multicolumn{1}{c}{$i$ $[^{\circ}]$}\\
\hline
CD-31 3271 & 5.6204259(12) & 8488.468706(13) & 0.09298(32) & 0.098288(65) & 0.41841(32) & 0.8190(24) & 89.390(20) \\
HD 4875 & =13.635561$^1$  & 8738.6151(14) & 0.10149(49) & 0.06171(14) & 0.3250(29) & 0.7940(39) & 89.057(25) \\
HD 287990 & 12.937763(34) & 9177.7561(18) & 0.21669(83) & 0.07808(30) & 0.3790(69) & 0.8195(49) & 88.792(98) \\
TYC 8547-22-1 & 4.241822564(16) & 8867.7506448(43) & 0.11690(12) & 0.163910(95) & 0.35975(79) & 0.8056(12) & 86.201(13) \\
BD-08 1175 & 17.7032485(12) & 8479.970812(32) & 0.17668(34) & 0.052450(44) & 0.41732(30) & 0.8046(25) & 89.5341(91) \\
\hline
Name &
\multicolumn{1}{c}{$e\cos\omega$}& 
\multicolumn{1}{c}{$e\sin\omega$} &
\multicolumn{1}{c}{$\ell$$^{*}$} & 
\multicolumn{1}{c}{$K_1$ [km/s]} &
\multicolumn{1}{c}{$V_0$ [km/s]} &
\multicolumn{1}{c}{$K_2$$^{**}$ [km/s]} & \\
\hline
CD-31~3271 & =0  & =0  & 0.014828(51) & 41.249(11) & 24.1332(97) & 100.141(74) & \\
HD~4875 & -0.055883(24) & -0.17227(46) & 0.00996(21) & 29.739(17) & -11.220(12) & 75.13(16) &\\
HD~287990 & -0.334412(23) & -0.0078(11) & 0.0284(11) & 40.602(96) & 39.846(51) & 80.99(23) & \\
TYC~8547-22-1 & 0.000119(13) & 0.00828(44) & 0.013923(63) & 50.981(85) & 42.832(71) & 111.70(13) &\\
BD-08~1175 & -0.179472(13) & 0.13139(31) & 0.028337(51) & 33.462(15) & 108.749(15) & 62.17(12) & \\
\hline
\noalign{\smallskip}
\multicolumn{8}{p{0.8\textwidth}}{{\bf Notes:} $^{*}$ Derived parameter. $^{**}$ $K_2$ values were determined separately through the cross-correlation method described in Sect.~\ref{sec:spectroscopy}. $^1$ From \cite{maxted:2023}.}
\end{tabular}

\end{table*}

\subsection{Mass and radius measurements}
\label{sec:fundamental_par}
With our combined analysis of light curves, radial velocities, and cross-correlation, we have sufficient information to derive $M_{\star}$, $R_{\star}$ as well as surface gravity $\log{g}$ and mean density $\rho_{\star}$. The masses can be calculated from the orbital parameters. Using the total mass ($M_1+M_2$) of the system, the stellar radii can be determined from Kepler's third law combined with the fractional radius and radius ratio provided by {\sc jktebop}. With both radius and mass obtained for the components we can then determine the surface gravity $\log g$ and mean stellar density $\rho_{\star}$. 

\begin{figure*}
	\includegraphics[width=0.8\textwidth]{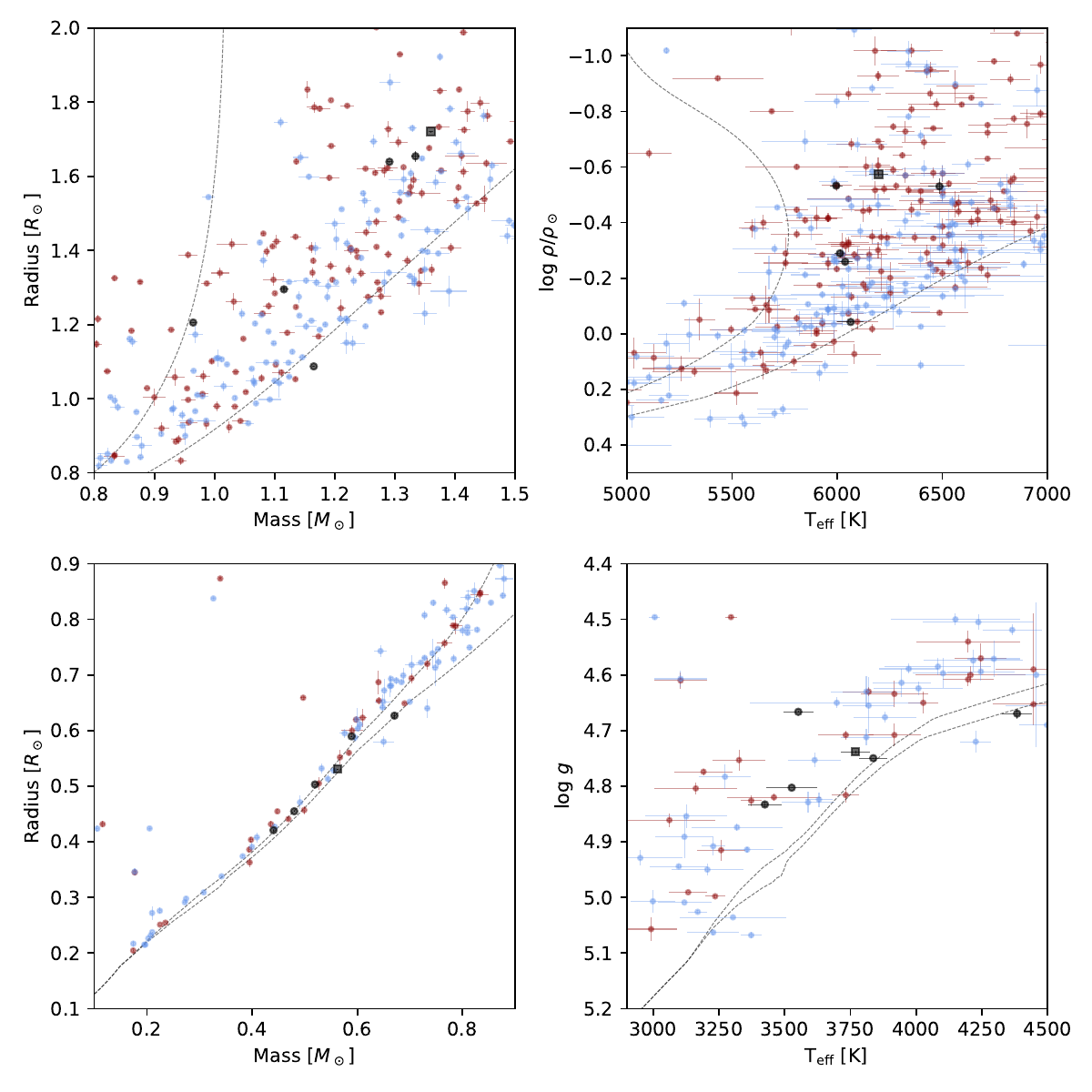}
    \caption{
    Properties of the primary stars (upper panels) and secondary stars (lower panels) obtained from our analysis. CD-27 2812, from \citet{adshead:2026} is shown as a square. Other detached eclipsing binaries from DEBCat \citep{southworth:2015} are included as red and blue points for the primary and secondary, respectively. Isochrones with solar metallicity, taken from the MIST grid of stellar models \citep{choi:2016}, at ages of 1 and 10~Gyr are shown as dashed lines.}
    \label{fig:massradius}
\end{figure*}

The mass, radius, surface gravity and mean density estimates for each component are given in Table~\ref{tab:massradius} and plotted in Fig.~\ref{fig:massradius} together with other binary stars collected from DEBCat \citep{southworth:2015}.

\begin{table*}
\centering
\caption{Fundamental stellar parameters for the binary systems in this study.}
\label{tab:massradius}
 \begin{tabular}{lcrrrrrr} 
 \hline
 \multicolumn{2}{l}{Star} &
 \multicolumn{1}{l}{$M [M_{\odot}]$} &
 \multicolumn{1}{l}{$R [R_{\odot}]$} &
 \multicolumn{1}{l}{$\log{g}$} &
 \multicolumn{1}{l}{$\rho/\rho_{\odot}$} &
 \multicolumn{1}{l}{$T_{{\rm eff},1}$ [K]}  &
 \multicolumn{1}{l}{$\log{L/L_{\odot}}$}
 \\
 \hline
\multirow{2}{*}{CD-31~3271} & A & 1.1651(25) & 1.0872(10) & 4.43100(67) & 0.9066(18) & 6064(62) & 0.157(18)\\
 & B & 0.48001(62) & 0.45490(51) & 4.80269(81) & 5.099(15) & 3527(83) & -1.542(41)\\
\multirow{2}{*}{HD~4875} & A & 1.1150(66) & 1.2950(59) & 4.2600(38) & 0.5135(65) & 6013(25) & 0.2910(60)\\
 & B & 0.4405(15) & 0.4209(21) & 4.8329(43) & 5.909(89) & 3425(41) & -1.662(18)\\
\multirow{2}{*}{HD~287990} & A & 1.3431(95) & 1.654(15) & 4.1268(75) & 0.2952(76) & 6486(55) & 0.627(14)\\
 & B & 0.6732(38) & 0.6269(61) & 4.6692(88) & 2.715(83) & 4384(29) & -0.935(14)\\
\multirow{2}{*}{TYC~8547-22-1} & A & 1.2908(43) & 1.6391(25) & 4.1189(10) & 0.29314(95) & 5995(30) & 0.4917(89)\\
 & B & 0.5891(20) & 0.58966(95) & 4.6663(13) & 2.873(12) & 3552(29) & -1.306(15)\\
\multirow{2}{*}{BD-08~1175} & A & 0.9644(44) & 1.2055(19) & 4.2592(12) & 0.5505(15) & 6038(34) & 0.238(10)\\
 & B & 0.5195(15) & 0.50307(86) & 4.74964(86) & 4.081(13) & 3837(20) & -1.3092(90)\\
 \hline
\multicolumn{8}{l}{{\bf Notes:} Secondary temperatures are subject to additional systematic uncertainties of $\sim50$~K.}
\end{tabular}
\end{table*}

\subsection{Direct measurement of the stellar effective temperature}
\label{sec:teb}
The bolometric luminosity $L$ of a star with Rosseland radius $R$ is given by 
\begin{equation}
    L=4\pi R^2 \sigma_{\rm SB} {\rm T}_{\rm eff}^4,
\end{equation}
where $\sigma_{\rm SB}$ is the Stefan-Boltzmann constant. For a binary star at distance $d$ (with a parallax $\varpi=1/d$), the flux, corrected for interstellar extinction, is
\begin{equation}
    f_{0,b}= f_{0,1}+f_{0,2}=\frac{\sigma_{\rm SB}}{4}\left[\theta_1^2{\rm T}_{\rm eff,1}^4 + \theta_2^2{\rm T}_{\rm eff,2}^4\right],
\end{equation}
with the angular diameter $\theta_i=2R_i\varpi$. To extract effective temperatures of both targets, we use {\sc 
teb}\footnote{\url{https://github.com/nmiller95/teb}} \citep{miller:2020,maxted:2025}.  The code utilises {\sc emcee} \citep{foremanmackey:2013} to obtain posteriors of $T_{{\rm eff},i}$, $\theta_i$, E(B--V), and systematic errors given observed magnitudes and flux ratios.

The radii obtained from the light curve analysis in Sect.~\ref{sec:light curve} is not the Rosseland radius as stellar disks appears slightly larger than what their fundamental radius would imply. This is because the flux close to the limb originates from above the photospheric radius. For high precision radii, this could impact our effective temperature measurements. We test the impact of this discrepancy on our measurements by using the Solar correction of 0.333\,Mm from \cite{haberreiter:2008}. To adjust the value to different stellar atmospheres, we scale it by the atmospheric pressure scale height of our targets. For this sample of stars, correcting for the Rosseland radius is a marginal effect, causing a $<0.1\%$ radius effect on all targets. The change in radius caused by this correction is also lower than the radius uncertainty for all targets.

For each star, {\sc teb} calculates synthetic photometry by using a model spectral energy distribution (SED). Optionally, this SED may be multiplied by a distortion function that is a linear superposition of Legendre polynomials. The advantage of this approach is that the distortion function allows the shape of the SED to adapt to the observed data while still including realistic effects such as stellar absorption features. This means that we are more sensitive to the integrated flux, rather than the specific shape of the model SED. The SEDs used in this work are computed using linear interpolation and BT-Settl model atmospheres \citep{allard:2013}. The code allows an arbitrary number of distortion coefficients in the fit. To select a suitable number, we considered the BIC and AIC values produced by {\sc teb} for different distortion coefficients. We find that four distortion coefficients per stellar component is generally the preferred number of distortion coefficients for our targets. For this reason, we choose to use four distortion coefficients per component for the fit, unless specified otherwise in Sect.~\ref{sec:individual}.

We include a prior on the interstellar reddening $\rm E(B-V)$ determined from the interstellar Na~I doublet lines at 589-590\,nm following the semi-empirical relationship from \cite{maxted:2025a}. The spectra for this analysis is produced by stacking the individual HARPS spectra, each divided by the mean spectra from Sect.~\ref{sec:spectroscopy}, in the barycentric rest frame. For the stars where no significant absorption lines were detected, we set $\rm E(B-V)$ to zero, with a $1\sigma$ upper limit of 0.014 as this is a typical uncertainty from the dataset used in \cite{maxted:2025a}. For specific values, see details of individual stars in Sect.~\ref{sec:individual}.

{\sc teb} automatically extracts magnitudes from a range of surveys. For this work, we use magnitudes from GALEX \citep{bianchi:2014}, Gaia DR3 \citep{Gaiacollaboration:2023}, 2MASS \citep{skrutskie:2006}, WISE \citep{cutri:2012}, TYCHO \citep{hoeg:2000}, and Sky Mapper \citep{onken:2024}. The parallaxes are also automatically retrieved from Gaia and corrected using the zero-point corrections from \cite{flynn:2022}. We also provide the flux ratios $l$ of the TESS bands as provided in Table~\ref{tab:lcfits}. Besides the intrinsic uncertainties provided, {\sc teb} also includes additional exponential priors on the uncertainties for magnitudes and flux ratio. These can be used to account for systematic effects in the input data.

\begin{figure}
    \centering
    \includegraphics[width=\linewidth]{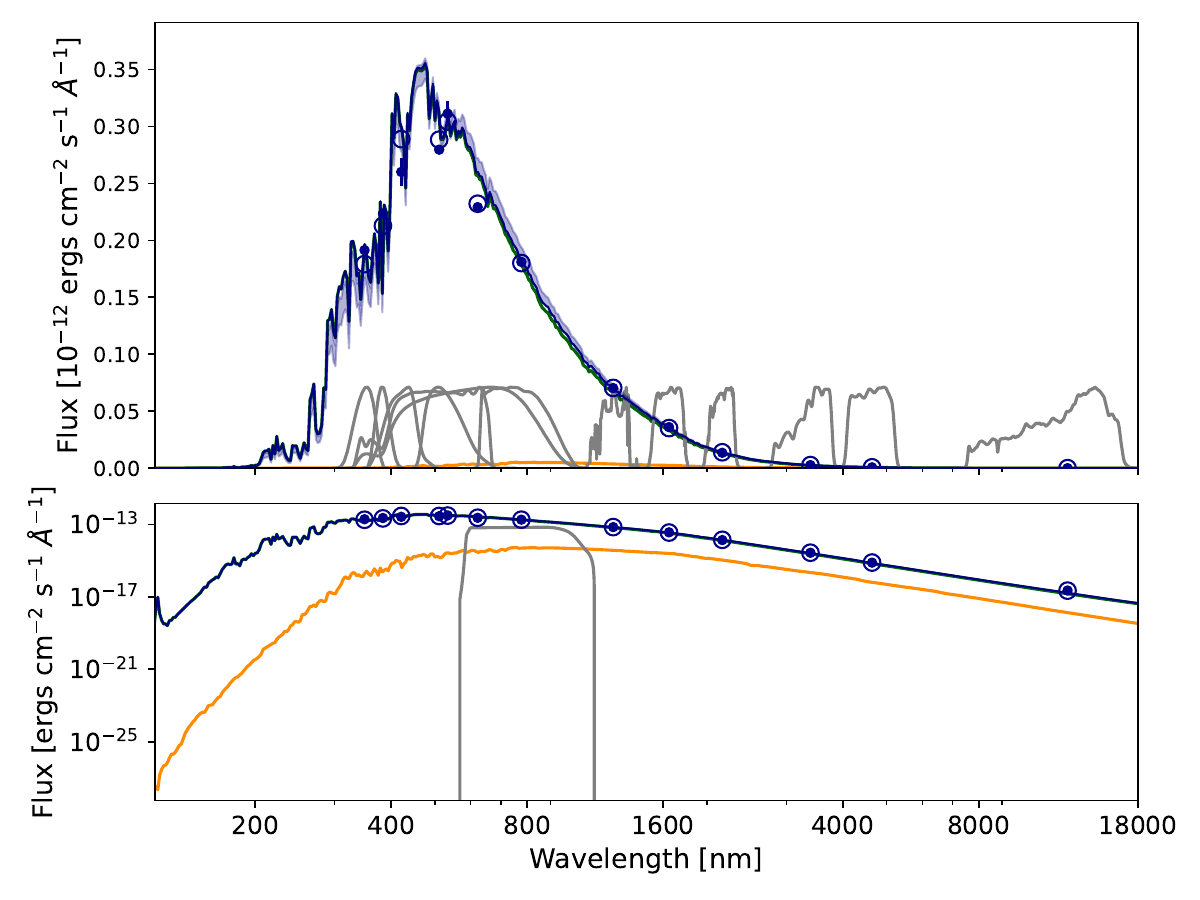}
    \caption{Upper panel: SED of BD-08~1175. The best-fit SED is plotted as a dark blue and the mean SED $\pm 1-\sigma$ is plotted as a filled region. The observed fluxes are plotted as points with error bars and predicted fluxes for the best-fit SED integrated over the response functions shown are plotted with open circles. The SEDs of the individual components are shown as teal and orange lines, respectively. Lower panel: Same as the upper panel, but on a logarithmic scale. Filters used to measure flux ratios are also plotted here.}
    \label{fig:SED_plot}
\end{figure}

Many of the magnitudes have uncertainties that are typically larger than the expected flux contribution of the secondary. For this reason, it is challenging to extract precise information for the secondary as only one flux ratio is available. At the same time, the secondary can vary quite substantially without significantly affecting the effective temperature of the primary component. To estimate effective temperatures of the secondary components, we first run {\sc teb} to get an idea of the primary effective temperature. From that temperature, we use the average surface brightness ratio\footnote{Note that $J_0$ in Table~\ref{tab:lcfits} is the flux ratio at the disk centre, unaffected by limb darkening.} $J = \ell/k^2$ to provide a more accurate estimate of the secondary effective temperature as discussed in \cite{maxted:2023}. With this constraint, we rerun {\sc teb} with the new temperature estimate and a stronger constraint on the flux ratio prior. Comparing the two sets of runs, the primary effective temperatures are consistent within their uncertainties. While some of the secondary temperatures show significant deviation between the two runs, the second set of runs is better aligned with the flux ratios measured from TESS. A note about this procedure is that the constraint could introduce systematic uncertainties as the average surface brightness ratio will depend on the stellar model used to estimate the the average surface brightness ratio. \cite{maxted:2025} found a variation of $\sim50$~K due to this uncertainty. For this reason, it is likely that our secondary temperatures are subject to additional uncertainties of about 50~K. The temperatures of both components, from the second run, are given in Table~\ref{tab:massradius}. An example of the fit, and the resulting SED, for BD-08~1175 is shown in Fig.~\ref{fig:SED_plot}.

\section{Notes on individual systems}
\label{sec:individual}
\label{sec:notes}
\subsection{HD 4875}

When comparing astrometry from Hipparcos and Gaia astrometry the proper motion of HD 4875 has been found to vary \citep{frankowski:2007,brandt:2021}. \cite{kervella:2019} interpreted this variation in proper motion as a companion with a mass of $\approx 40\,M_{\rm Jup}$. We find support for this interaction, as we were unable to fit the radial velocities and TESS light curves with the same ephemerides. Specifically, we saw a offset between the two datasets of about 0.007 phase units. This is likely due to the fact that the photometric and spectroscopic observations of HD~4875 are obtained with a separation of a few years. To mitigate this issue, we fix $P$ to the value from \cite{maxted:2023}, and introduce an offset in $T_0$ for the spectroscopic data when using {\sc jktebop} (effectively fitting two different values for the two datasets). The $T_0$ in Table~\ref{tab:lcfits} refers to the TESS data, but we advise caution using the linear ephemerides provided for HD 4875 due to the likely presence of a third component that influences the orbital motion. We also verified that this approach had no significant effect on the resulting parameters compared to fitting the spectroscopic and photometric data separately.
We also note a slight third light contribution for HD~4875, while only significant to $1\sigma$, we include a third light component into our fit, resulting in a value of $l_3=0.024\pm0.019$.

As HD~4875 were only observed on two nights with HARPS, the outlier removal when producing the mean spectra is not possible. While we detected no strong interstellar absorption in the spectra. Any non-stellar features are for this reason unlikely to be properly removed in the average spectra. 

We find good agreement between our obtained masses and radii when compared with estimates from \cite{maxted:2023}. Both of our effective temperatures are hotter, by about 250 and 100~K for the primary and secondary, respectively. This difference is comparable to the 200\,K uncertainties from \cite{maxted:2023}.

\subsection{HD 287990}
Similar to HD~4875, there is a third light contribution with a $1\sigma$ significance. Our value mean value is $l_3 = -0.052\pm0.040$. This negative third light value may be an instrumental effect, e.g. inaccurate scattered light correction,  or due to star spots, or a combination of these effects.

The reddening estimate of HD 287990 using \ion{Na}{I} doublet lines is $\rm E(B-V)\sim0.01$, this value is affected by the fact that the shape of the \ion{Na}{I} doublet lines are not represented by Gaussians. In addition, as only 3 HARPS spectra were observed, the outlier rejection approach when stacking the spectra fails to remove much of the interstellar absorption around the \ion{Na}{I} doublet lines. 

When determining the number of distortion coefficients most suitable for the temperature measurement using {\sc teb}, we found that the BIC and AIC preferred 3 coefficients for HD~287990 instead of the typical 4 found for our other targets. For this reason, we use 3 distortion coefficients for HD~287990. This choice should have a low impact on the resulting temperatures as initial testing did not indicate any substantial difference in effective temperatures between using 3 or 4 distortion coefficients.

\subsection{CD-27 2812}
This star was studied in \cite{adshead:2026}. As similar methods were used to study this star, we include a brief summary of the results here. Furthermore, its orbital period ($P=7.8$~days) falls in between our short ($<6$~days) and long ($>12$~days) period systems. This makes it an interesting point of comparison to our systems, for this reason, it is included in Fig.~\ref{fig:massradius}

The primary is an F9 star with fundamental parameters, $R_1=1.72\,R_{\odot}$, $M_1 = 1.36\,M_{\odot}$, and $T_{\rm eff,1} = 6197$\,K. The secondary has the parameters $R_2=0.53\,R_{\odot}$, $M_2=0.56\,M_{\odot}$, and $T_{\rm eff,2}=3770$\,K. Based on the best-fit age of the primary star, the radius of the secondary star was noted to be matching with predictions from the MIST stellar evolution model grid. This indicates that the star does not exhibit significant radius inflation, in contrast to what is commonly observed among M-dwarfs.

\subsection{CD-31 3271}

\cite{maxted:2023} reported star spot signatures in the TESS light curves with amplitudes of up to 2~per~cent. Although this could justify the use of a third light component, we found a value consistent with zero when fitting the light curve. This is an intermediate value compared to the range of values reported by \cite{maxted:2023}. We find an increase in $T_{\rm eff,1}$  $\approx 200$~K compared to the values reported in \cite{maxted:2023}.
 
\subsection{TYC 8547-22-1}
\cite{maxted:2023} reported a clear star spot signal in the light curve with an amplitude of about 0.5~per~cent. similar to CD$-$31 3271, we included $\ell_3$ as a free parameter in the light curve analysis. Our obtained third light value was $l_3=0.022\pm0.005$. 

We find that the primary is about 0.1\,$M_\odot$ more massive than predicted by \cite{maxted:2023} using $K_1$ values from Gaia spectra. We also find temperatures about 100~K hotter than previously predicted on both components. The spectroscopic analysis also shows that TYC~8547-22-1 has the largest Ti\,I offset with respect to the GBS v3 reference abundances. This star also shows the strongest line broadening in our sample, which makes the selected Ti\,I lines more difficult to analyse. In particular, the Ti\,I features appear more blended and less well defined than in the other stars. Therefore, the low Ti\,I abundance obtained for this star is likely related to the quality of the selected Ti\,I lines, rather than to the M-dwarf correction itself.

\subsection{BD-08 1175}
When determining effective temperatures with {\sc teb}, we found that the magnitudes from Pan-STARRS \citep{tonry:2018} provided anomalous values compared to other sources. Checking the data manually, we found no correlation between the observation and any primary eclipses (offset is too large to be caused by any secondary eclipse alone). We find two sources within $2''$ but $>1''$ away from the expected position of BD-08 1175. The data is also flagged as poor-quality due to a lack of good measurements. For these reasons we elect to exclude the Pan-STARRS data from the fitting of BD-08-1175 when using {\sc teb}.

From the \ion{Na}{I} doublet lines, BD-08~1175 also exhibit an interstellar reddening at $\rm E(B-V)\sim0.03$. In contrast to HD~287990, the interstellar \ion{Na}{I} doublet lines are well described by a Gaussian profile for BD-08~1175. 

\section{Discussion}
\label{sec:discussion}
\subsection{Tidal interaction}
\label{sec:tide}

\begin{table}
    \centering
    \caption{$v\sin i$ estimates and tidal synchronisation timescales}
    \begin{tabular}{lcrrrr}
        \hline
        Star &  & \multicolumn{1}{c}{$v_{\rm sync}\sin i$ } & \multicolumn{1}{c}{$v\sin i$} & \multicolumn{1}{c}{P$_{\rm rot}$} & \multicolumn{1}{c}{$\tau_{Q}$}\\
         & & \multicolumn{1}{c}{[km\,s$^{-1}$]} & \multicolumn{1}{c}{[km\,s$^{-1}$]} & \multicolumn{1}{c}{[d]} & \multicolumn{1}{c}{[Gy]} \\
        \hline
        \noalign{\smallskip}
        \multirow{2}{*}{CD-27 2872$^*$} & A &  12.6 & 7.0 & $12.4 \pm 0.9$ & 2.4 \\
                       & B &   3.9 & 3.0 & $ 9.0 \pm 2.1$ & 5.7 \\
        \multirow{2}{*}{CD-31~3271}     & A & 9.8 & $8.7 \pm 0.5$ & $6.3 \pm 0.4$ & 2.6 \\
                       & B & 4.1 & $3.9 \pm 0.7$ & $5.9 \pm 1.1$ & 2.5 \\
        \multirow{2}{*}{HD~4875} & A & 5.9 & 3.0\parbox{0pt}{$^\dag$}    & $>$22          & 130 \\
               & B & 1.9 & $1.3 \pm 0.7$ & $16.4 \pm 8.8$ & 240 \\
        \multirow{2}{*}{HD~287990} & A & 8.1 & $2.9 \pm 0.3$ & $28.7 \pm 3.0$ & 42 \\
                  & B & 3.1 & $1.6 \pm 0.7$ & $19.8 \pm 8.7$ & 98\\
        \multirow{2}{*}{TYC~8547-22-1} & A & 19.9 & $19.4 \pm 0.2$ & $4.27 \pm 0.04$ & 0.2 \\
                      & B &  7.2 &  $7.1 \pm 0.7$ &  $4.2 \pm 0.4$ & 0.4\\
        \multirow{2}{*}{BD-08~1175} & A & 4.3 & 2.4\parbox{0pt}{$^\dag$} & $> 25$        &  340 \\
                   & B & 1.8 &         $3.2\pm 0.7$   & $8.0\pm 1.7$ &  740\\
        \noalign{\smallskip}
        \hline
    \multicolumn{6}{l}{$^{*}$ Parameters derived or taken from \cite{adshead:2026}. } \\
    \multicolumn{6}{l}{$^\dag$ Upper limits using $v_{\rm mac}$ from \cite{doyle:2014}.}
    \end{tabular}

    \label{tab:vsini2}
\end{table}
Close-in binary systems often exhibit circularisation of the orbit and tidal locking, which synchronises the rotational period with the orbital period. Before the orbit has circularised, binary stars may still exhibit pseudosynchronisation \citep{hut:1981}, where the components rotation rate aligns with the orbital angular velocity near the periastron. We aim to evaluate to what extent our targets exhibit tidal interactions by comparing rotation rates predicted by pseudosynchronisation model from \cite{hut:1981} with spectroscopic estimates of $v\sin{i}$. Understanding the extent of tidal interaction is important, as such interactions may impact the long-term evolution of the star, and by extension affecting their suitability as benchmark stars when comparing them with single stars. 

The spectroscopic values of $v\sin{i}$ for the primary is determined from the HARPS spectra as described in Sect.~\ref{sec:optical_spectroscopy}. For some targets, the obtained $v\sin i$ values were low compared to the macroturbulent velocities, this could introduce substantial systematic errors in the estimated rotation rate depending on choice of macroturbulence. For these targets (marked with a $\dag$ in Table~\ref{tab:vsini2}), we also estimate the macrotubulent velocity using the empirical relationship from \cite{doyle:2014} and use the higher of the two as an estimated upper limit on the rotation rate. This is then used to calculate a lower limit of the rotation rate. 

To obtain a spectroscopic estimate of the $v \sin i$ of the secondary components, we compare the CCF functions from Sect.~\ref{sec:spectroscopy}, with CCF functions generated from synthetic spectra with rotational broadening applied. We use the same resolving power as our NIRPS observations ($R=70\,000$, see \ref{sec:NIRPS}). Based on 3D-simulations of M dwarf surfaces \citep{wende:2009} that finds no significant velocity fluctuations in M dwarf atmospheres, we set the macroturbulent velocities to 0 when applying broadening to the synthetic spectra. We then find the value of $v\sin i$ where the observed and synthetic CCF have a matching shape. From the $v\sin i$ values of both components, we calculate expected rotation periods to compare with the orbital period from Sect.~\ref{sec:light curve}. The resulting values can be seen in Table~\ref{tab:vsini2}. 

We find that for the targets with a short orbital period ($P < 6$ days), the agreement between the spectroscopic and pseudosynchronous estimates are quite good. For the other objects ($P > 12$ days), the agreement is less consistent. Looking at an intermediate case, CD-27~2812 from \cite{adshead:2026} (period of $\sim8$\,days), we find that both components are rotating slightly slower than predicted by by pseudosynchronisation.

The likely explanation for this is that the timescales where these objects are expected to synchronise their orbital rotation is much shorter for the short period objects, giving the star time to synchronise to the theoretical predictions. We calculated synchronisation timescales (see Table~\ref{tab:vsini2}) using relationships from \cite{barker:2020} assuming a rotation period of $P_{\rm rot} = 1.1P_{\rm orb}$, changing the assumed rotational rate does change the expected synchronisation timescale, but not significantly. We find that the stars with shorter tidal synchronisation timescales are the short period targets with $v\sin{i}$ values well aligned between pseudosynchronisation and CCF measurements. For the slow rotators, we seem to find that the secondary components are more aligned with pseudosynchronisation, even if they should have longer synchronisation timescales. Finding rotation periods, independent from spectroscopic estimates of $v\sin{i}$ for these targets would be useful to verify this trend.

An issue at lower rotation rates is that the $v\sin i$ is no longer the dominant source of broadening. As discussed in Sect.~\ref{sec:optical_spectroscopy} some amount of systematic effects are introduced from our selection of macroturbulent broadening. Similarly, a low $v\sin{i}$ likely diminishes our ability to accurately measure the $v\sin i$ from the CCF, particularly if there are other artifacts such as imperfectly removed tellurics or signal from the primary still present in the spectra. It is also possible that the orbital and rotation axis are not aligned, which could result in a $v\sin i$ lower than expected. 

From the results it seems that pseudosynchronous rotation produces reliable estimates of short period targets where tidal synchronisation should be effective. This is in line with results from \cite{torres:2010}, which showed that rotation rates of stars in binaries with fractional radii above $\sim0.1$ agree well with pseudosyncronous theory. \cite{sethi:2026} also finds that most binaries with short periods are synchronised. In addition, they find that larger mass ratio tends to improve tidal synchronisation. For mass ratios $>0.4$, synchronisation appears effective up to orbital periods of about 6--7 days. While they only looked at the primary stars of the system, we find that their results are also in line with the rotational measurements of the M-dwarf components. This indicates that similar relationships should hold for the low-mass companions.

\subsection{Stellar ages}
\label{sec:ages}

\begin{figure}
    \centering
    \includegraphics[width=\linewidth]{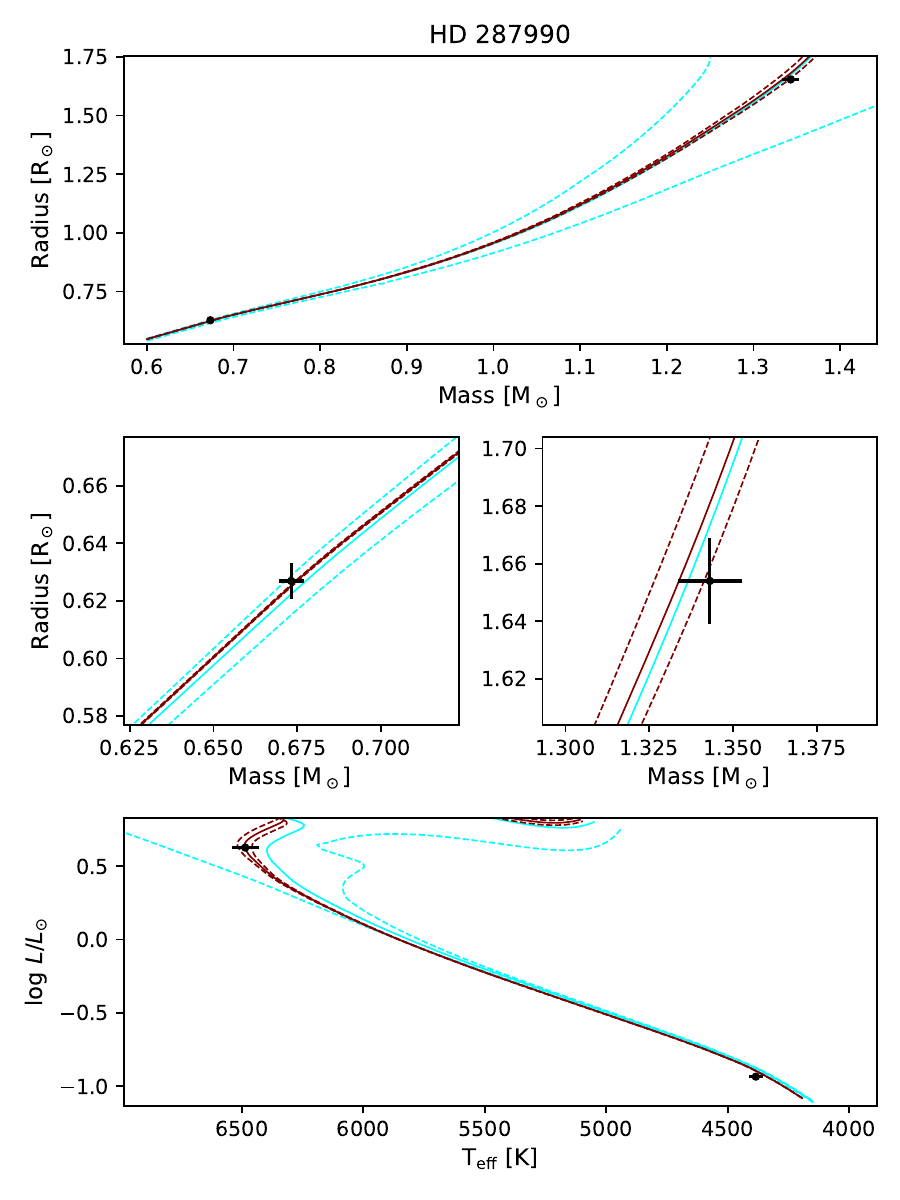}
    \caption{The best-fit isochrones for HD~287990 from {\sc bagemass}. The primary isochrone is shown in dark red and the isochrone for the secondary is shown in teal. Dashed lines represents $1\sigma$ uncertainties in age. Black points with error-bars are the parameters of the stellar components. \textit{Top:} Isochrones in mass-radius space. \textit{Middle:} Zoom in of the top row close to each component. \textit{Bottom:} Same isochrones plotted in a Hertzsprung-Russel diagram.}
    \label{fig:bagemass}
\end{figure}

With a collection of accurately characterised stars, we also estimate their ages using {\sc bagemass} \citep[for details, see][]{maxted:2015}. {\sc bagemass} uses a grid of stellar evolution tracks calculated using the {\sc garstec} stellar evolution code \citep[see][for details]{weiss:2008,serenelli:2013} to estimate mass ($M$), initial metallicity ([Fe/H]$_i$), and age ($\tau$) posterior distributions given information on effective temperature (T$_{\rm eff}$), luminosity (L), surface metallicity ([Fe/H]), and observed density ($\rho$) using MCMC sampling. Due to our well constrained mass from Sect.~\ref{sec:light curve}, we apply a Gaussian prior on mass using our obtained uncertainties. For [Fe/H]$_i$ and $\tau$, we use uninformative uniform priors over the entire range covered by the grid ($-0.75<$ [Fe/H] $<0.55$, $0<\tau<17.5$\,Gyr). Our input parameters use uncertainties from Sect.~\ref{sec:METHODS}. As the stellar evolution grid used in {\sc bagemass} has a mass range of 0.6 to 2\,M$_\odot$ we are generally unable to determine the ages of the secondary components. The ages presented here are therefore determined from the primary component only. The exception to this is HD~287990, with a secondary component mass of $0.67$\,M$_\odot$ we can also use {\sc bagemass} on the secondary to estimate how well the ages of the two components align. The resulting isochrones for HD~287990 can be seen in Fig.~\ref{fig:bagemass}. Our estimated ages for all stars can be seen in Table~\ref{tab:bagemass}, the typical age for our sample is a few Gyr with CD-31~3271 standing out as only a few hundred Myr and BD-08~1175 being more evolved at above 7\,Gyr.

For HD~287990, we find that the age estimates are generally consistent, although the uncertainty in age for the secondary is significantly larger. As the fit is separate, there is also some variation in parameters such as initial metallicity ($\Delta$[Fe/H]$_{\rm ini} \approx 0.07$). For the other stars, we can not make a direct comparison due to the mass limit of {\sc bagemass}. However, we can take the estimated age of the primary and use isochrones from MIST \citep{choi:2016} to evaluate how well the secondary components agree with the primary age. The isochrones are calculated by taking best matching metallicity (steps of 0.25~dex) in the grid of MIST isochrones and interpolating to the age determined from the primary. The MIST isochrones and comparison to the {\sc bagemass} can be seen in Fig.~\ref{fig:isos}. 

While some care should be taken in comparing different evolutionary models with slightly different metallicities, we find that some secondary components exhibit a larger radius than predicted by the MIST isochrones at the age given by the primary. Specifically, the secondary components in the short period binaries show larger deviations between the observed and expected radii compared to the longer period binaries. As the connection between magnetic activity and rotation rates are well established \citep[e.g.][]{vidotto:2014,reiners:2022} this radius inflation could be driven by enhanced magnetic activity in the short-period binaries due to tidal interactions. This is supported by our estimated rotation rates from Sect.~\ref{sec:tide} which show larger rotation rates for the M-dwarf components in our short period binaries.

\begin{table}
    \centering
    \caption{Ages determined for the binary components with $M_\star>0.6M_\odot$}
    \label{tab:bagemass}
    \begin{tabular}{ll}
       \hline
       Star & Age [Gyr] \\
       \hline
        CD-31~3271 & $0.3 \pm 0.1$ \\
        HD~4875 & $4.6 \pm 0.2$\\
        HD~287990 A &  $2.3 \pm 0.1$ \\
        HD~287990 B &  $3.2 \pm 1.6$\\
        TYC~8547-22-1 & $3.6 \pm 0.1$\\
        BD-08~1175 & $7.4 \pm 0.2$\\
       \hline
    \end{tabular}
    
\end{table}

\section{Conclusions}
\label{sec:conc}
In this work, we have characterised a group of 5 EBLM systems. Using high-quality spectroscopic and photometric data we obtained accurate fundamental parameters for both components in our targets. The primary stars will be useful benchmark systems for verifying automatic pipelines for future large-scale surveys as their spectroscopic signal shows negligible influence from the secondary components. The secondary components also represent a small sample of accurately characterised M-dwarfs. This is a very interesting sample to test models on stellar structure for low-mass stars, as accurate characterisation of single M-dwarfs is difficult. Improvements in characterisation of the M-dwarf components would involve flux ratio measurements in multiple bands in addition to our TESS measurements. This would allow the effective temperatures to be better constrained using {\sc teb}. One unfortunate limitation of these EBLM systems, due to the low flux ratio, is that no spectra of the individual M-dwarf components are currently available. Specific follow-up on the properties of the secondary, such as independent chemical composition or stellar activity, will therefore remain a challenge. The latter is particularly interesting, as some of the more rapidly rotating M-dwarf companions studied in this work seems to exhibit radius inflation, possibly coupled to stellar magnetism.

\section*{Acknowledgements}

A.H. and P.F.L.M acknowledges support from UK Science and Technology Facilities Council (STFC) by the grant UKRI1193. I.H.A. acknowledges financial support from FONDECYT Regular Grant 1231057. C.A.G. acknowledges support from Agencia Nacional de Investigación y Desarrollo (ANID) through FONDECYT Regular 1262342.

This research made use of Lightkurve, a Python package for Kepler and TESS data analysis \citep{lightkurvecollaboration:2018}. Beyond the packages mentioned in the text, this work have also utilised the following Python packages: NumPy \citep{Harris:2020}, SciPy \citep{virtanen:2020}, Astropy \citep{AstropyCollaboration:2013}, Matplotlib \citep{Hunter:2007}

This paper includes data collected by the TESS mission obtained from the MAST data archive at the Space Telescope Science Institute (STScI). Funding for the TESS mission is provided by the NASA Explorer Program. STScwe is operated by the Association of Universities for Research in Astronomy, Inc., under NASA contract NAS 5–26555.

This research has made use of the VizieR catalogue access tool, CDS, Strasbourg, France (DOwe : 10.26093/cds/vizier). 

\section*{Data Availability}

The data underlying this article are available from the following sources: MAST data archive at the Space Telescope Science Institute (STScI) (\url{https://archive.stsci.edu}); ESO archives (\url{https://archive.eso.org/scienceportal/home}).







\appendix



\section{Radial velocities}
\begin{table*}
    \caption{Observed radial velocities}
    \label{tab:rvs}
\begin{tabular}{@{}rrlrrl}
        \\
        \hline
        ${\rm BJD} -2450000$ & \multicolumn{1}{c}{$V_1$ [km/s]} & Instrument & 
        ${\rm BJD} -2450000$ & \multicolumn{1}{c}{$V_1$} [km/s] & Instrument\\
        \hline
         \hline
        \multicolumn{3}{@{}l}{CD-31~3271}   &      \multicolumn{3}{@{}l}{TYC~8547-22-1} \\
         10226.85939 &$-15.235\pm0.034$&HARPS  &   10221.84306 &$-7.430\pm0.080$&HARPS\\
         10226.86031 &$-15.190\pm0.066$&NIRPS  &   10221.84377 &$-7.251\pm0.080$&NIRPS\\
         10271.72592 &$-16.232\pm0.065$&NIRPS  &   10248.78119 &$66.099\pm0.081$&HARPS\\
         10276.67795 &$-10.996\pm0.064$&NIRPS  &   10248.78231 &$66.354\pm0.093$&NIRPS\\
         10301.68264 &$50.771\pm0.035$&HARPS  &    10302.64672 &$-7.554\pm0.079$&HARPS\\
         10301.68343 &$50.778\pm0.066$&NIRPS  &    10302.64755 &$-7.385\pm0.077$&NIRPS\\
         10367.66361 &$-9.149\pm0.064$&NIRPS  &    10381.58992 &$76.992\pm0.081$&HARPS\\
         10369.57489 &$62.531\pm0.064$&NIRPS  &    10381.59034 &$77.025\pm0.084$&NIRPS\\
         10383.57827 &$-13.439\pm0.062$&NIRPS  &   10383.55314 &$1.212\pm0.080$&HARPS\\
         10383.58206 &$-13.492\pm0.034$&HARPS  &   10383.55406 &$1.170\pm0.077$&NIRPS\\
         10386.62851 &$64.866\pm0.035$&HARPS  &    10385.64102 &$86.246\pm0.080$&HARPS\\
         10386.62951 &$64.886\pm0.063$&NIRPS  &    10385.64162 &$86.239\pm0.083$&NIRPS\\
         10387.61014 &$48.369\pm0.034$&HARPS  &    10387.58224 &$-5.802\pm0.079$&HARPS\\
         10387.61107 &$48.434\pm0.065$&NIRPS  &    10387.58311 &$-5.824\pm0.079$&NIRPS\\
         10398.56940 &$57.527\pm0.035$&HARPS  &    10398.59771 &$74.648\pm0.080$&HARPS\\
         10398.57021 &$57.517\pm0.066$&NIRPS  &    10398.59850 &$74.644\pm0.094$&NIRPS\\
         10400.60216 &$-15.962\pm0.034$&HARPS  &   \multicolumn{3}{@{}l}{BD-08~1175} \\
         10400.60302 &$-15.897\pm0.062$&NIRPS  &  10187.87883 &$116.446\pm0.045$&NIRPS\\
         \multicolumn{3}{@{}l}{HD~4875}   &       10213.80412 &$118.372\pm0.046$&NIRPS\\
         10128.85803 &$-7.166\pm0.047$&NIRPS  &    10221.86731 &$104.233\pm0.021$&HARPS\\
         10130.82595 &$-26.613\pm0.047$&NIRPS  &   10221.86828 &$104.405\pm0.048$&NIRPS\\
         10180.89254 &$12.691\pm0.046$&NIRPS  &    10251.84466 &$77.909\pm0.021$&HARPS\\
         10185.73337 &$-30.086\pm0.046$&NIRPS  &   10251.84557 &$78.011\pm0.044$&NIRPS\\
         10222.70028 &$6.664\pm0.020$&HARPS  &     10300.79964 &$131.535\pm0.021$&HARPS\\
         10222.70081 &$6.736\pm0.044$&NIRPS  &     10300.80054 &$131.589\pm0.044$&NIRPS\\
         10226.68578 &$-30.600\pm0.019$&HARPS  &   10366.65645 &$127.617\pm0.022$&HARPS\\
         10226.68644 &$-30.510\pm0.045$&NIRPS  &   10366.65654 &$127.658\pm0.047$&NIRPS\\
         10270.65718 &$-37.464\pm0.045$&NIRPS  &   10378.53500 &$77.110\pm0.021$&HARPS\\
         10274.56048 &$16.354\pm0.043$&NIRPS  &    10378.53605 &$77.217\pm0.041$&NIRPS\\
         10275.55887 &$16.001\pm0.044$&NIRPS  &    10379.55617 &$87.439\pm0.021$&HARPS\\
         10276.62034 &$11.043\pm0.044$&NIRPS  &    10379.55683 &$87.532\pm0.043$&NIRPS\\
         10281.61428 &$-33.971\pm0.046$&NIRPS  &   10385.55003 &$132.770\pm0.020$&HARPS\\
         10478.89884 &$15.745\pm0.045$&NIRPS  &    10385.55109 &$132.885\pm0.043$&NIRPS\\
         10499.91385 &$-35.043\pm0.045$&NIRPS  & \\
         10530.72752 &$-22.198\pm0.049$&NIRPS  &  \\
         10546.85874 &$14.744\pm0.046$&NIRPS  &   \\
         \multicolumn{3}{@{}l}{HD~287990}   &  \\
         10247.76041 &$65.943\pm0.027$&HARPS  &  \\
         10247.76435 &$66.157\pm0.061$&NIRPS  &  \\
         10271.76956 &$65.562\pm0.062$&NIRPS  &  \\
         10272.78176 &$67.103\pm0.061$&NIRPS  &  \\
         10300.79048 &$60.593\pm0.028$&HARPS  &  \\
         10300.79168 &$60.792\pm0.063$&NIRPS  &  \\
         10368.56520 &$12.701\pm0.058$&NIRPS  &  \\
         10369.55835 &$-11.405\pm0.051$&NIRPS  &  \\
         10386.52305 &$55.092\pm0.028$&HARPS  &  \\
         10386.52697 &$55.277\pm0.058$&NIRPS  &  \\
\hline

\end{tabular}
\end{table*}

\section{Abundances}

\begin{figure*}
    \centering
    \includegraphics[width=1\textwidth]{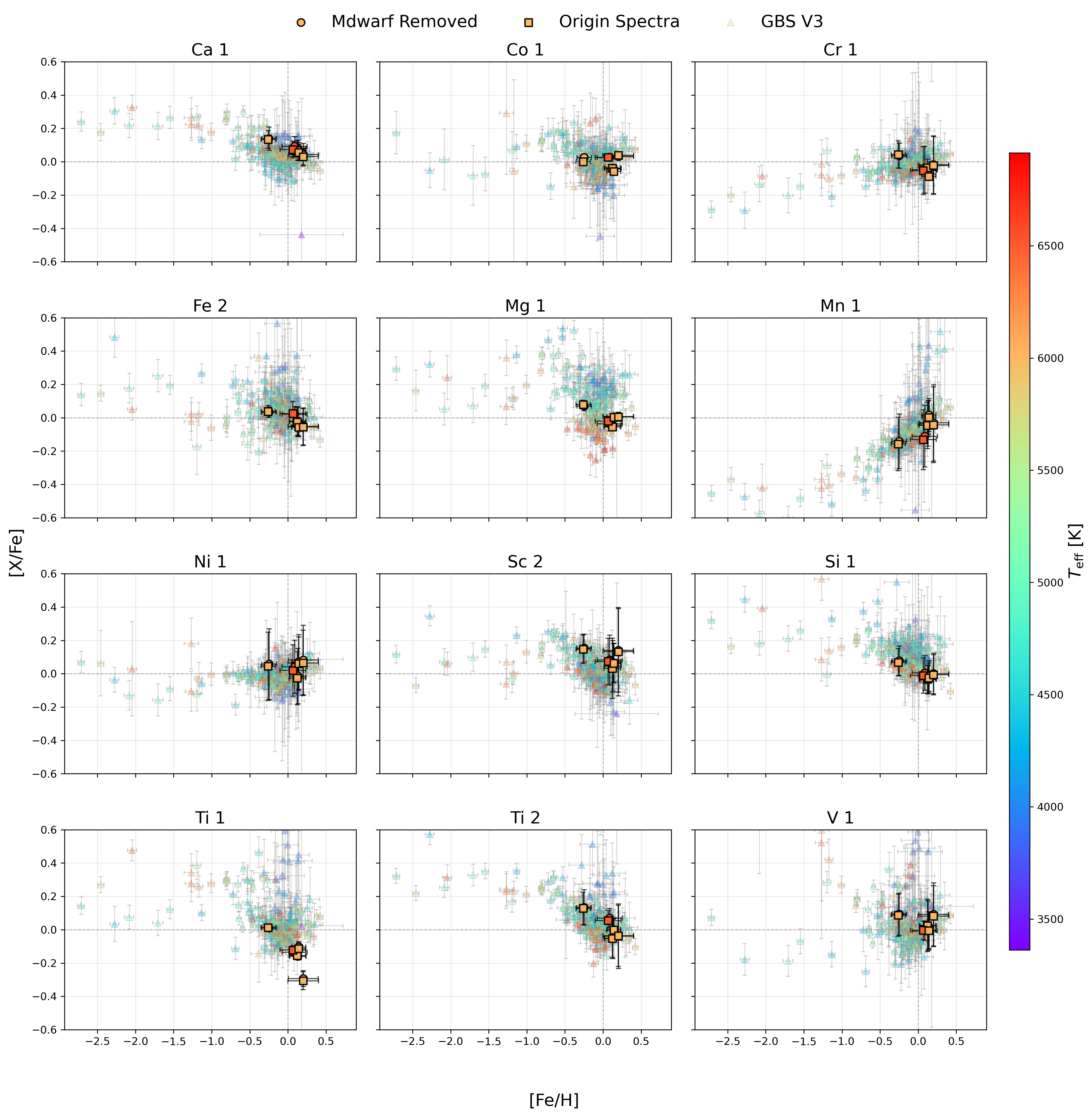}
    \caption{
    Comparison between the abundances derived in this work and the GBS v3 reference abundances. Each panel shows $[X/Fe]$ as a function of $[Fe/H]$ for a different chemical species. The reference GBS v3 abundances are shown as triangles and are color-coded by effective temperature. Our abundances derived from the original spectra are shown as squares, while those obtained from the M-dwarf corrected spectra are shown as circles. The uncertainties on the individual abundances were estimated from the line-to-line scatter, computed as the standard deviation of the abundances derived from the individual spectral lines used for each element.}
    \label{gbsv3_abundances}
\end{figure*}

\begin{table*}
    \centering
    \caption{Chemical abundances measured with the HARPS spectra of each star as described in Sect~\ref{sec:optical_spectroscopy}.}
    \label{tab:gbsv3_abundances}
    \begin{tabular}{lrrrrr}
        \hline
        [X/Fe]  & BD-08~1175 & CD-31~3271 & HD~287990 & HD~4875 & TYC~8547-22-1\\
        \hline
        Ca~1 & 0.134(53) & 0.067(58) & 0.073(51) & 0.055(48) & 0.030(50)\\
        Co~1 & $-$0.000(00) & $-$0.038(00) & 0.027(00) & $-$0.058(00) & 0.039(00)\\
        Cr~1 & 0.043(81) & $-$0.035(50) & $-$0.05(14) & $-$0.089(23) & $-$0.02(17)\\
        Fe~2 & 0.037(36) & $-$0.025(85) & 0.026(70) & $-$0.057(55) & $-$0.06(11)\\
        Mg~1 & 0.079(25) & $-$0.055(19) & $-$0.020(25) & 0.003(21) & 0.005(25)\\
        Mn~1 & $-$0.16(16) & $-$0.05(14) & $-$0.13(18) & 0.00(10) & $-$0.04(23)\\
        Na~1 & 0.1914(93) & 0.147(19) & 0.269(21) & 0.185(28) & 0.361(21)\\
        Ni~1 & 0.05(20) & $-$0.03(16) & 0.02(15) & 0.06(16) & 0.06(20)\\
        Sc~2 & 0.147(86) & 0.03(18) & 0.07(14) & 0.06(12) & 0.13(27)\\
        Si~1 & 0.068(82) & $-$0.033(83) & $-$0.01(10) & $-$0.025(89) & $-$0.01(12)\\
        Ti~1 & 0.012(14) & $-$0.155(20) & $-$0.122(26) & $-$0.113(34) & $-$0.306(54)\\
        Ti~2 & 0.129(95) & $-$0.05(11) & 0.055(60) & $-$0.002(66) & $-$0.04(18)\\
        V~1 & 0.09(13) & 0.03(16) & $-$0.004(55) & $-$0.01(11) & 0.08(18)\\
        Y~2 & $-$0.29(11) & $-$0.224(98) & $-$0.117(92) & $-$0.13(10) & $-$0.69(31)\\
        \hline
    \end{tabular}
    
\end{table*}

\section{Binary Isochrones}
\begin{figure*}
    \centering
    \includegraphics[width=0.4\textwidth]{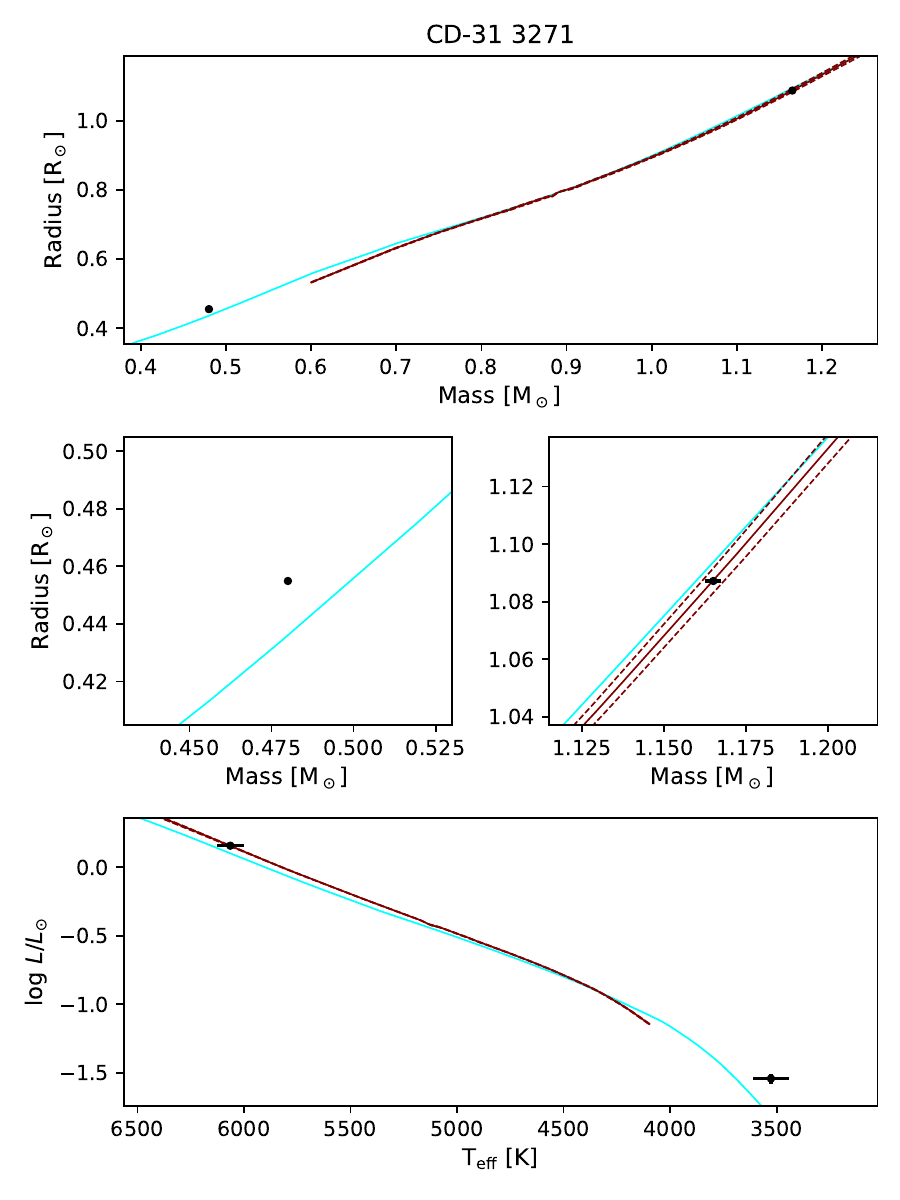}
    \includegraphics[width=0.4\textwidth]{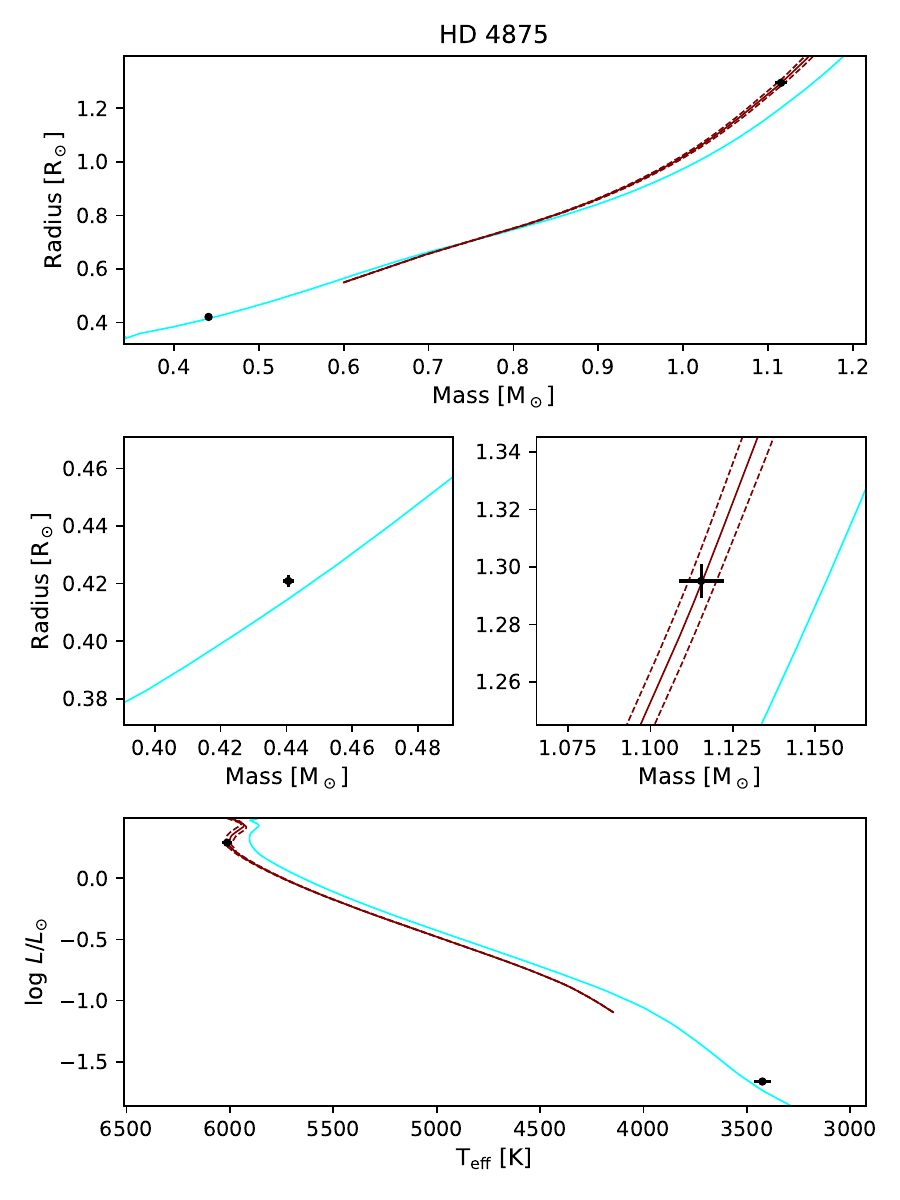}
    \includegraphics[width=0.4\textwidth]{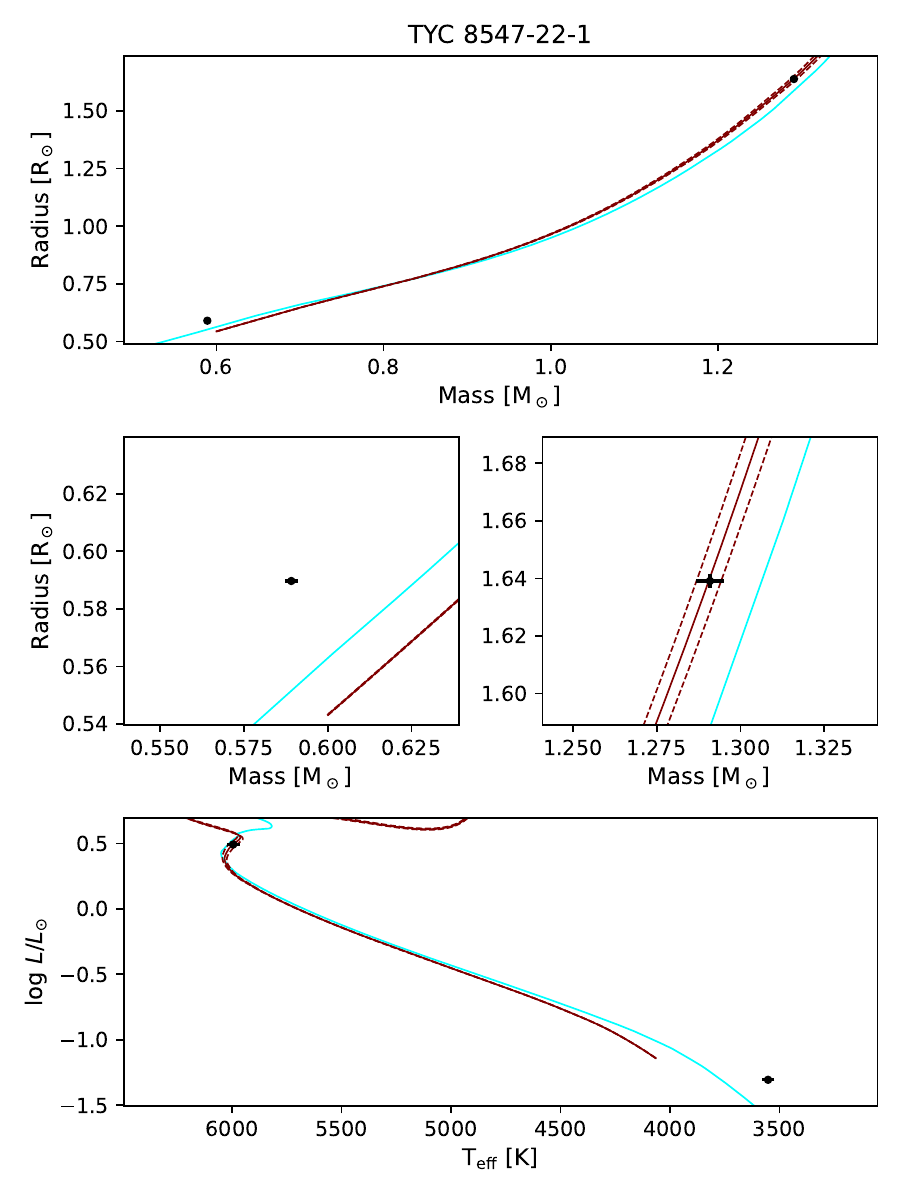}
    \includegraphics[width=0.4\textwidth]{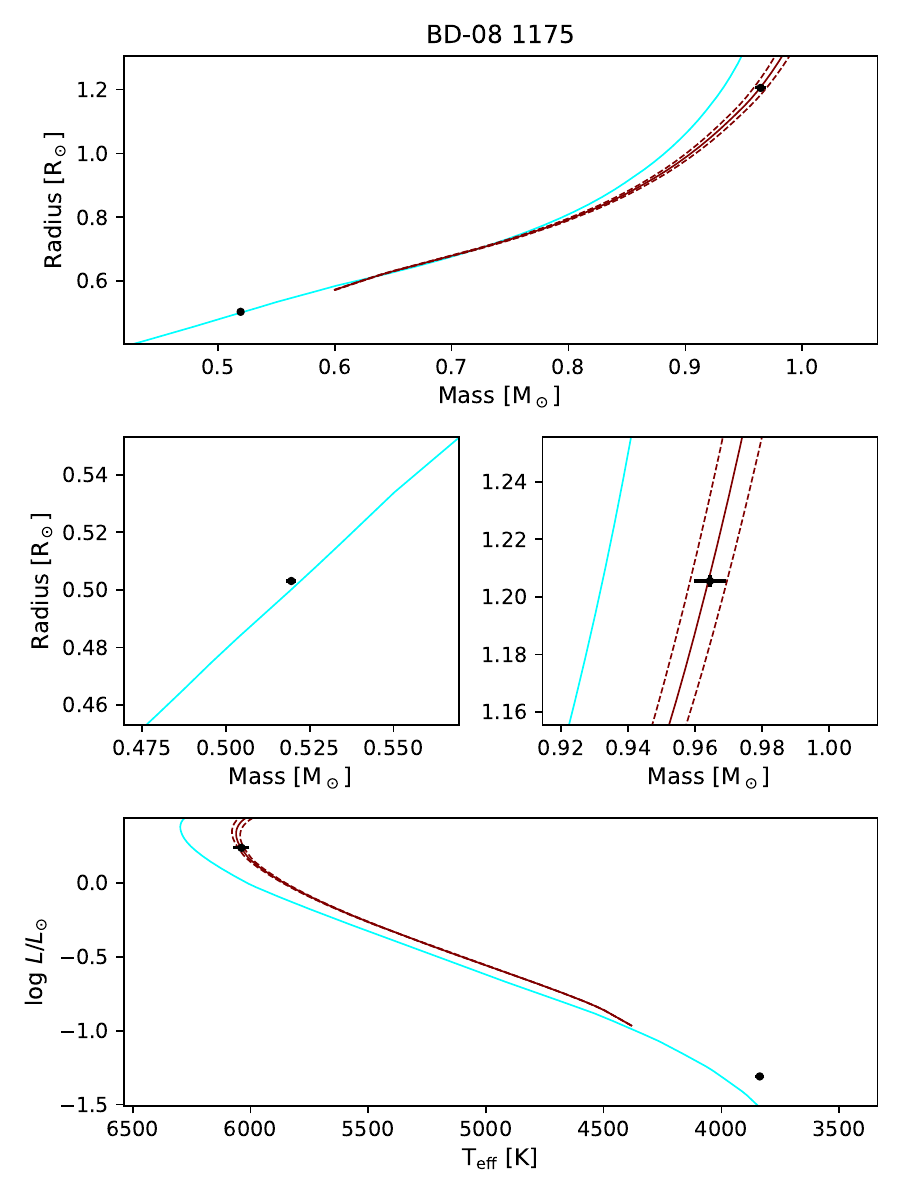}
    \caption{Isochrones for binary systems. The individual figures are similar to Fig.~\ref{fig:bagemass} except that the teal isochrone is taken from MIST.}
    \label{fig:isos}
\end{figure*}

\bsp	
\label{lastpage}

\end{document}